 \documentclass[prd,aps,showpacs,preprintnumbers,amsmath,amssymb,nofootinbib,preprintnumbers]{revtex4}
\usepackage{graphicx}
\usepackage{epsf}
\usepackage{amsmath}
\usepackage{amssymb,latexsym}
\usepackage{epstopdf}
\usepackage{bm}
\usepackage{color}
\usepackage{tabularx}
\usepackage{enumitem}
\usepackage{float}
\usepackage{array,booktabs}
\usepackage{footnote}
\usepackage{threeparttable}
\usepackage{graphicx}
\usepackage{ulem}
\usepackage{hyperref}
\usepackage{amssymb,epsf}
\usepackage{latexsym}
\usepackage{epstopdf}
\usepackage{epsfig}
\usepackage{subfigure}
\usepackage{upgreek}

\begin{document}
     \title{Gravitational Radiation in Generalized Brans Dicke Theory: Compact Binary Systems

     }
     \author{
        S. Mahmoudi$^{1,2}$\footnote{email address: S.mahmoudi@shirazu.ac.ir} and
        S. H. Hendi$^{1,2,3}$\footnote{email address: hendi@shirazu.ac.ir}
        }
     \affiliation{
        $^1$Department of Physics, School of Science, Shiraz University, Shiraz 71454, Iran \\
        $^2$Biruni Observatory, School of Science, Shiraz University, Shiraz 71454, Iran \\
        $^3$Canadian Quantum Research Center 204-3002 32 Ave Vernon, BC V1T 2L7 Canada }

     \begin{abstract}

    This paper investigates the generation and properties of
    gravitational radiation within the framework of Generalized
    Brans-Dicke (GBD) theory, with a specific emphasis on its
    manifestation in compact binary systems. The study begins by
    examining the weak field equations in GBD theory, laying the
    foundation for subsequent analyses. Solutions to the linearized
    field equations for point particles are then presented, providing
    insights into the behavior of gravitational fields in the context
    of GBD theory. The equation of motion of point mass is explored,
    shedding light on the dynamics of compact binary systems within
    the GBD framework. The primary focus of this study lies in the
    comprehensive exploration of gravitational radiation generated by
    compact binaries. The energy momentum tensor and the associated
    gravitational wave (GW) radiation power in GBD theory are
    investigated, elucidating the relationship between these
    fundamental concepts. Furthermore, detailed calculations are
    provided for the GW radiation power originating from both tensor
    fields and scalar fields. Based on our calculations,  both scalar
    fields contribute to GW radiation by producing dipole radiation.
    We also study the period derivative of compact binaries in this
    theory. By comparing with the observational data of  the orbital
    period derivative of the quasicircular white dwarf-neutron star
    binary PSR J1012+5307, we put bounds on the two parameters of the
    theory: the Brans-Dicke coupling parameter $\omega_{0}$ and the
    mass of geometrical scalar field $m_f$, \textcolor{black}{resulting
    	in a lower bound $\omega_{0}>6.09723\times10^6$ for a massless BD
    	scalar field and the geometrical field whose mass is smaller than
    	$10^{-29} \text{GeV}$. The obtained bound on $\omega_0$ is two
    	orders of magnitude stricter than those derived from solar system
    	data.} Finally, we find the phase shift that GWs experience in the
    frequency domain during their propagation. These calculations
    offer a quantitative understanding of the contributions from
    different components within the GBD theory, facilitating a deeper
    comprehension of the overall behavior and characteristics of
    gravitational radiation.
     \end{abstract}

     \maketitle

\section{Introduction}
Einstein's theory of General Relativity (GR), since its inception in 1916, is the most successful classical theory
of gravity whose predictions were examined with extremely high precision and it perfectly explained all related phenomena in the weak-field and low energy  regimes \cite{Weinberg:1972kfs}.
 However, there are some long-standing puzzles in GR, like understanding the nature of dark matter \cite{Sahni:2004ai} and resolving the singularity problem \cite{Joshi:2013xoa}, due to which GR gets modified in the classical framework which deviates from its original version in ultraviolet and/or infrared energy scales.
 Another unanswered issue is related to realizing the observed late time cosmic acceleration \cite{SupernovaCosmologyProject:1997zqe, SupernovaCosmologyProject:1998vns, SDSS:2005xqv, WMAP:2006bqn}, leading to introducing the dark energy, i.e. a new exotic form
 of stress-energy as the supplementary material in our universe, to explain this phenomenon within the framework of GR. Nonetheless, alternative theories of gravity are also worth trying to find an appropriate explanation for this observation. In this regard, an enormous amount of activities tried to address this issue based on the possibility that the acceleration may be caused by modification of gravity at large scales \cite{DeFelice:2010aj, Nojiri:2010wj, Nojiri:2017ncd, Faraoni:2008mf}. Moreover, many other studies were focused on the context of scalar-tensor gravity, mainly based on the assumption of cosmological scalar fields rather than postulating a mysterious form of dark energy \cite{Amendola:1999dr, Bento:2001yv, Nojiri:2003ft, Nojiri:2005pu, Elizalde:2004mq, Faraoni:2006sr, Catena:2006bd, Elizalde:2008yf}.
 The two simplest modifications of GR in these directions are Brans-Dicke (BD) gravity and $f(R)$ theory. \par
  BD theory, as a special case of scalar-tensor theory, is obtained by substituting a time modifying gravitational constant $G(t)$ by means of a scalar field $\phi$, which interacts with the geometry by a non-minimal coupling ($\omega_{BD}$) with gravity. According to BD theory, the gravitational interaction is mediated by both the curvature of the space as well as a scalar field which is assumed to be the reciprocal of the dynamical gravitational constant, i.e., $G(t)=\frac{1}{\phi(t)}$. Indeed, in BD theory a non-linear self interaction of the scalar field allows the possibility of understanding the late time accelerated expansion of the universe while recovering the GR behavior at early cosmological epochs. The large value of $\phi$ describes the fast expansion of the universe and is found by a recent study in cosmology,
  including the redshift and distance-luminosity connection of type Ia Supernovae \cite{SupernovaSearchTeam:1998fmf}. The evidence for various cosmic
  concerns, including the late behavior of the universe, cosmic acceleration, and the inflation issue, etc. are also
  supported by the BD theory \cite{Banerjee:2000mj}.\par
     For observational and theoretical motivations, the original BD theory has been extended in several ways, like adding a potential term to the original BD theory \cite{BD-potential}, assuming the time-varying coupling constant $\omega$ \cite{BD-omegat1,BD-omegat2}, etc. In this regard, authors in Ref. \cite{Lu:2018nvu} proposed a different way to extend the original BD theory by replacing a function $f(R)$ instead of the Ricci scalar $R$ in the BD action, enabling a more comprehensive description of gravitational phenomena  \cite{Lu:2018nvu, Lu:2019qgk}. Regarding the importance of this modified gravity (abbreviate as GBD), there are some interesting properties compared to other theories. For instance, it is renowned that the $f(R)$ theory is equivalent to the BD theory with a potential for choosing a specific value of the BD parameter $\omega=0$. This is while the specific choice $\omega=0$ for the BD parameter is considered quite exceptional, and as a result, the corresponding absence of the kinetic term for the field is incomprehensible. However, for the GBD theory, the situation is similar to the coupled scalar-fields model, and both fields in the GBD have a non-disappeared dynamical effect \cite{Lu:2018nvu}. Besides, the GBD theory tends to investigate physics from the geometric point of view, while the BD theory (with $\omega=0$) with a potential has a tendency to study physical problems from the viewpoint of matter \cite{Lu:2018nvu}. Furthermore, the GBD theory can be viewed as a particular case of the more complicated $f(R,\phi)$ theory \cite{Hwang:1990re, Hwang:1996xh, Hwang:2005hb} and since the simpler theory is usually more favored by researchers in physics, here, we proceed to explore the GBD theory.\par
   
   In recent years, the LIGO and Virgo collaborations have detected several GWs from binary systems, providing an excellent opportunity to perform both weak field tests of the propagation of GWs, as well as tests of the strong field regime of compact binary sources \cite{Kobakhidze:2016cqh, Baibhav:2019rsa, Chia:2020dye, Kobakhidze:2017mru, Arunasalam:2017ajm, LIGOScientific:2016lio, GuerraChaves:2019foa, Caldwell:2022qsj, Odintsov:2022cbm, Qiao:2022mln, Ireland:2023avg, Ng:2022agi, Ghoshal:2023fhh}. Compact binary systems are excellent laboratories to test theories of gravity in the strong field regime. Since the coalescence of a compact binary system can produce
   strong gravitational fields, the GW observations allow us to test GR and modified gravity in the highly dynamic and strong field regime. The first indirect evidence of GW radiation was obtained from the observations of orbital decay of the Hulse-Taylor binary pulsar whose orbital period loss confirms Einstein's GR \cite{Peters:1963ux} to $0.1$ percent
   accuracy \cite{Weisberg:2016jye}. In fact, in order to better
   understand gravity and fundamental physics from these
   observations, it is important to clarify the corresponding
   predictions from GR and alternative theories of gravity
   \cite{Yunes:2013dva, Clifton:2011jh}. Therefore, the study of gravitational radiation in alternative theories of gravity has become an important issue. \par
   
   In this paper, we continue the work that has been done by authors in \cite{Lu:2019qgk}. Our primary objective is to analytically investigate the characteristics of GW emission predicted by the GBD theory. More precisely,  We will pursue two goals in the present paper: First, the rate of energy loss and orbital period decay of compact binary systems are derived from the aforementioned GBD theory to put constraints on the parameter space of the theory using the observational data of PSR J1012+5307 binary system. Next, we explore the influence of varying GBD parameters
   on the observed GW signals, particularly the phase function of GWs. \par
   
   The paper is organized as follows. In Section \ref{weak field}, we rederive
   the gravity field equation and the weak-field expansion of the field equations. Section \ref{point mass} is devoted to studying point particles in linearized GBD theory: Solutions of the linearized field equations for point mass particles are found in section \ref{point particle}. Section \ref{EOM point mass} begins to demonstrate the new results. In section \ref{EOM point mass}, we investigate the equation of motion of point particles using the
   Einstein-Infeld-Hoffmann (EIH) method. Next, we study the gravitational radiation generated by compact binaries in section \ref{GW radiation}. To this end, we first obtain the energy momentum tensor (by employing the perturbed field equations approach) and the gravitational radiation power formula in GBD theory in section \ref{Energy momentum}. Then, the contributions of the tensor field and two scalar fields are calculated in sections \ref{GW radiation1} and \ref{GW radiation2}, respectively. Using the obtained results, we try to put constraints on GBD theory from the observational data related to the orbital period decay of a particular compact binary.
   Section \ref{phase function} is devoted to finding the frequency domain phase shift that GWs experience during propagation within the GBD framework. Section \ref{conclusion} concludes and points to possible directions for future research. \par
   
   Before we proceed, it should be mentioned that we set the units so that $c=\hbar=1$. We do not set  $G$ equal to 1, since the effective gravitational constant depends on the background value of the scalar field, which will vary over the history of the universe.


\section{ Weak field equations in GBD theory}\label{weak field}

 In this section, we study the weak field limit of GBD theory by employing a function $f(R)$ to replace the Ricci scalar $R$  in the original BD action. Hence, the action of the system takes the following form \cite{Lu:2019qgk}
 \begin{equation}
 	A=\frac{1}{16\pi}\int \sqrt{-g}\Big[\phi f(R)- \frac{\omega(\phi)}{\phi}\partial _\mu \phi \partial ^\mu \phi+M(\phi)\Big]d^4x\,+\,A_{m}\Big[g_{\mu\,\nu}, \Psi_{m}\Big].\label{action}
 \end{equation}
 where $g\equiv\,\text{det} g_{\mu\,\nu}$ and $\omega(\phi)$ is the coupling function which is responsible for the spontaneous scalarization phenomenon \cite{Damour:1993hw}. The cosmological function $M(\phi)$ can provide the effective cosmological constant and the mass of scalar field. $A_{m}$ denotes the matter action and $\Psi_{m}$ is
 the matter fields.

 \par
 In the following, we will assume that  $\omega$ is the coupling constant, i.e. $\omega=\omega_{0}$, and the scalar field $\phi$ is massless ($M(\phi)=0$). Thus, the action (\ref{action}) reduces to
 \begin{equation}
 	A=\frac{1}{16\pi}\int \sqrt{-g}\Big[\phi f(R)- \frac{\omega_{0}}{\phi}\partial _\mu \phi \partial ^\mu \phi\Big]d^4x\, +A_{m}.\label{action2}
 \end{equation}
 Varying the action (\ref{action2}), one can obtain the gravitational field equation and  the BD scalar field equation as follows
 \begin{eqnarray}
 	\mathcal{G}_{\mu\nu}=\phi \left[f_{R}R_{\mu\nu }-\frac{1}{2}f(R)g_{\mu \nu }\right]- (\nabla_\mu\nabla_\nu -g_{\mu\nu}\Box)(\phi f_{R})+ \frac{\omega_{0}}{\phi}\Big[\frac{1}{2}g_{\mu\nu}\partial_\sigma\phi\partial^\sigma\phi
 	-\partial_\mu\phi\partial_\nu\phi\Big]= 8\pi T_{\mu \nu },\nonumber\\
 	\label{gravitational equation}
 \end{eqnarray}
 \begin{equation}
 	f(R)+2\omega_{0}\frac{\Box \phi}{\phi} -\frac{\omega_{0}}{\phi^{2}}\partial _\mu \phi \partial ^\mu \phi=-16\pi\,\frac{dT}{d\phi},\label{scalar-eq}
 \end{equation}
 where $f_{R}\equiv \partial f/\partial R$, $\nabla _\mu $ is the covariant derivative associated with the Levi-Civita connection of the metric,  $\Box \equiv \nabla ^\mu \nabla _\mu $, and $T$ is trace of the stress-energy tensor, $T_{\mu \nu }=\frac{-2}{\sqrt{-g}}\frac{\delta A_m}{\delta g^{\mu \nu }}$.
 The trace of Eq. (\ref{gravitational equation}) is
 \begin{eqnarray}
 	f_{R}R-2f(R)+\frac{3\Box (\phi f_{R})}{\phi}+\frac{\omega_{0}}{\phi^{2}}\partial_\mu\phi\partial^\mu\phi
 	= \frac{8\pi T}{\phi}.\label{trace}
 \end{eqnarray}
 Combining Eqs. (\ref{scalar-eq}) and (\ref{trace}), we get
 \begin{eqnarray}
 	\Box \phi-\frac{\partial_{\mu}\phi\partial ^{\mu}\phi}{4\phi}=\frac{1}{4\omega_{0}}\Big[8\pi (T-4\,\phi\,\,\frac{d T}{d\phi}) -\phi R f_{R}-3\,\Box (\phi f_{R})\Big].\label{equation of phi}
 \end{eqnarray}
 Now, we try to obtain the weak field limit of the above equations. The weak field approximation is described by the
 perturbation of flat space-time, i.e tensor field  $h_{\mu\nu}$, defined as
 \begin{eqnarray}
 	g_{\mu\nu}=\eta_{\mu\nu}+h_{\mu\nu},\label{weak metric}
 \end{eqnarray}
 where $\eta_{\mu\nu}$ is the Minkowski metric, and $h_{\mu\nu}$ denotes a small deviation with respect to the flat spacetime, i.e. $|h_{\mu\nu}|\ll 1$.  The trace of the metric perturbation $h$ is given by $h=\eta^{\mu\nu}h_{\mu\nu}$. According to Eqs.(\ref{trace}) and (\ref{equation of phi}), the GBD theory could be considered as two scalar-fields theory, i.e. the BD field and the effectively geometrical field $f_{R}=\Phi$. For these two scalar fields, the weak-field approximations can be expressed as
 \begin{align}
 	\phi=\phi_{0}+\varphi,\label{weak-phi}\,\,\,\,\\
 	\Phi=\Phi_{0}+\delta\Phi,\label{weak-Phi}
 \end{align}
 with $|\varphi|\ll \phi_{0}$ and $|\delta\Phi|\ll \Phi_{0}$. In fact, the linearized theory will
 be analyzed with a little perturbation of the background,
 which is assumed to be given by a near Minkowski background, i.e. a Minkowski background plus $\phi = \phi_0$ and $\Phi=\Phi_0$ \cite{Corda:2010zza}.\par
 In order to derive the linearized equations, one should specify the function of $f(R)$ into the field equations \eqref{gravitational equation} and \eqref{equation of phi}. We will assume that $f(R)$ is analytic
 around a certain value $R = R_0$ so that it can be expressed as a power series as follows
 \begin{equation}\label{analytic fR}
 	f(R)=\sum_{n=0}^{\infty}\frac{f^n(R_0)}{n!}(R-R_0)^{n}.
 \end{equation}
 Since we wish to calculate the perturbation around the flat space-time, we require $f(R)$ to be analytic around $R_{0} = 0$ and thus the above equation will reduce to
 \begin{equation}\label{fR}
 	f(R)=a_0+a_1\,R+\,\frac{a_2}{2!}\,R^{2}+\frac{a_3}{3!}\,R^{3}+\cdots.
 \end{equation}
 To link it to the GR, we will set $a_1=1$. Moreover, considering a uniform flat space-time, i.e. $R = 0$, and using Eq. \eqref{trace}, one can find $a_0=0$.
 By this choice of $f(R)$, the effectively geometrical field $f_{R}$ reads as
 \begin{eqnarray}\label{fprim}
 	f_{R}\equiv f'(R) = 1 + a_2 R\,+\,\frac{1}{2}\,a_3\,R^{2}+\cdots,
 \end{eqnarray}
 where by comparing it to Eq. \eqref{weak-Phi}, one can find that the value of the background field $\Phi_0$ should be fixed to unity.
 Therefore, by ignoring the second-order and the higher terms and using Eq. \eqref{fR}, the linearized equations of the gravitational field and both scalar fields can be expressed as
 \begin{align}
 	{R}_{\mu\nu}^{\left(1\right)}-\frac{{R}^{\left(1\right)}}{2}\eta_{\mu\nu}=
 	\partial_{\mu}\partial_{\nu}\,\delta\Phi^{(1)}+\partial_{\mu}\partial_{\nu}\frac{\varphi}{\phi_{0}}
 	-\eta_{\mu\nu}\Box_{\eta}\,\delta\Phi^{(1)}-\eta_{\mu\nu}\Box_{\eta}\frac{\varphi}{\phi_{0}}+\frac{8\pi T_{\mu\nu}}{\phi_{0}},\label{eq-weak-gravity}\\
 	\Box_{\eta}\varphi=\frac{3}{4\omega_{0}+3}[8\pi (T-4\,\phi\,\,\frac{d T}{d\phi})-\phi\,{R}^{\left(1\right)}\,-3\phi_{0}\Box_{\eta}\,\delta\Phi^{(1)}],\,\,\,\,\,\,\,\,\,\,\,\,\,\,\,\,\label{eq-weak-phi}\\
 	\Box_{\eta}\,\delta\Phi^{(1)}=\frac{{R}^{\left(1\right)}}{3}-\frac{\Box_{\eta}\varphi}{\varphi_{0}}+\frac{8\pi T}{3\phi_{0}},\,\,\,\,\,\,\,\,\,\,\,\,\,\,\,\,\,\,\,\,\,\,\,\,\,\,\,\,\,\,\,\,\,\,\,\,\,\,\,\,\,\,\,\,\,\,\,\,\,\,\,\,\,\,\,\,\,\,\,\,\,\,\,\,\,\,\,\,\,\,\,\,\,\,\,\,\,\,\,\,\,\,\,\,\,\,\,\,\,\,\label{eq-weak-Phi}
 \end{align}
 with  $\Box_{\eta}=\partial^{\sigma}\partial_{\sigma}$. Here, ${R}_{\mu\nu}^{\left(1\right)}$ and ${R}^{\left(1\right)}$  denote the linearized Ricci tensor and Ricci scalar, respectively. Also, $\delta\Phi^{(1)}$ refers to the first order perturbation of the scalar field $\Phi$ which based on Eq. \eqref{fprim} is equal to $a_{2}\,R^{(1)}$.\par
 
 Now we define a new tensor $\theta_{\mu\nu}$ as
 \begin{eqnarray}
 	\theta_{\mu\nu}=h_{\mu\nu}-\frac{1}{2}\eta_{\mu\nu}h-\eta_{\mu\nu}\frac{\varphi}{\phi_{0}}\,\,-\,\eta_{\mu\nu}h_{f},
 	\label{new tensor}
 \end{eqnarray}
 with $h_{f}\equiv \,\delta\Phi^{(1)}$
 and choose the following gauge condition
 \begin{eqnarray}
 	\partial^{\nu}\theta_{\mu\nu}=0.\label{Lorentz-gauge}
 \end{eqnarray}
 The expressions of the  linearized Ricci tensor and Ricci scalar will be obtained as
 \begin{align}
 	{R}_{\mu\nu}^{\left(1\right)}=\frac{1}{2}(2\partial_{\mu}\partial\nu h_{f}+2\partial\mu\partial\nu\frac{\varphi}{\phi_{0}}-\Box_{\eta}\theta_{\mu\nu}+\frac{\eta_{\mu\nu}}{2}\Box_{\eta}\theta+\,\eta_{\mu\nu}\Box_{\eta} h_{f}+\eta_{\mu\nu}\Box_{\eta}\frac{\varphi}{\phi_{0}}),\label{linear-Rmunu}\\
 	{R}^{\left(1\right)}=3\Box_{\eta} h_{f}+3\Box_{\eta}\frac{\varphi}{\phi_{0}}+\frac{\Box_{\eta}\theta}{2}.\,\,\,\,\,\,\,\,\,\,\,\,\,\,\,\,\,\,\,\,\,\,\,\,\,\,\,\,\,\,\,\,\,\,\,\,\,\,\,\,\,\,\,\,\,\,\,\,\,\,\,\,\,\,\,\,\,\,\,\,\,\,\,\,\,\,\,\,\,\,\,\,\,\,\,\,\,\,\,\,\,\,\,\,\,\,\,\,\,\,\,\,\,\,\,\,\,\,\,\,\,\,\,\,\,\,\,\,\,\,\label{linear-R}
 \end{align}
 where $\theta=\eta^{\mu\nu}\theta_{\mu\nu}=-h\,\,-\,4h_{f}-4\frac{\varphi}{\phi_{0}}$. Substituting Eqs. (\ref{linear-Rmunu}) and (\ref{linear-R}) into Eqs. (\ref{eq-weak-gravity})-(\ref{eq-weak-Phi}) leads to the following equations for the linearized gravitational field and two linearized scalar fields up to the first order perturbation
 \begin{eqnarray}
 	\Box_{\eta}\theta_{\mu\nu}&=&-\frac{16\pi T_{\mu\nu}}{\phi_{0}},\label{theta}\\ \Box_{\eta}\varphi&=&\frac{8\pi }{2\omega_{0}+3}(T^{\star}\,\,-\,\frac{3\,\phi_{0}\,}{8\pi} \Box_{\eta} h_{f}),\label{phi}\\
 	\Box_{\eta} h_{f}-m_{f}^2\,h_{f}&=&\,\frac{8\pi (T\omega_{0}+\phi\,\frac{dT}{d\phi})}{3\phi_{0}\,\omega_{0}},\label{Psi}
 \end{eqnarray}
 where $T^{\star}=T-2\,\phi\,\,\frac{d T}{d\phi}$,  $m_{f}^2\equiv\frac{{R}^{\left(1\right)}}{3\,\delta \Phi^{(1)}}\,\frac{2\omega_{0}+3}{2\omega_{0}}$ and $\omega_0 \neq -\frac{3}{2}$.  \textcolor{black}{In order to study the special case of $\omega_0=-\frac{3}{2}$, one can consider $T^{\star} = \frac{3\phi_0}{8\pi}\Box_{\eta} h_{f}$ . Although this special case is interesting from cosmological point of view, we will not delve into it here and leave it for future investigation.
 }\par
 
 Now, we are in a position to explore the possible solutions of the obtained linearized field equations.
 \section{Studying point particles in linearized GBD theory}\label{point mass}
 
 We start this part with the aim of solving the linearized field equations ($\ref{theta}$), ($\ref{phi}$), and ($\ref{Psi}$) to Newtonian order for point particles and then finding the equation of motion for a point mass. To this end, we follow the method described in \cite{Will:2018bme}.

 \subsection{{Solutions of the linearized field equations for point particles}}\label{point particle}

 The matter action for a system of point-like particles can be written as \cite{Weinberg:1972kfs}
 \begin{equation}
 	A_m=-\sum_a \int m_a(\phi)\ d\tau_a\label{mass action}
 \end{equation}
 where $m_a(\phi)$ and $\tau_a$ are, respectively, the inertial mass and the proper time of particle $a$ ($a^{th}$ particle) measured along its
 worldline $x^\lambda_a$. It is notable that according to the approach first proposed by Eardley \cite{Eardley}, since scalar fields can influence the self-gravity of compact objects, masses of particles are in general functions of scalar fields. Therefore, the stress-energy tensor and its trace take the form \cite{Weinberg:1972kfs}
 \begin{align}
 	T^{\mu\nu}=(-g)^{-\frac{1}{2}}\sum_am_a(\phi)
 	\frac{u^\mu u^\nu}{u^0}\delta^3({\bf{x}}-{\bf{x}}_a)\,, \\
 	T=g_{\mu\nu}T^{\mu\nu}=-(-g)^{-\frac{1}{2}}
 	\sum_a \frac{m_a(\phi)}{u^0}
 	\delta^3({\bf{x}}-{\bf{x}}_a)\,.\label{traceT}
 \end{align}
 where $u_a^\mu$ is the four velocity of particle $a$ and ${\bf x}_a$ is its position. Making use of the above information, we try to solve linearized field equations, obtained in the previous part, for point particles to Newtonian order:
 
 $\bullet$ The tensor field equation ($\ref{theta}$) reduces to
 \begin{equation}
 	\nabla^2\theta_{\mu\nu}=-\frac{16\pi}{\phi_0}T_{\mu\nu},
 \end{equation}
 with
 \begin{align}
 	\begin{split}
 		T_{00}=&\rho+O(\rho\,v^2),\\
 		T_{0i}=&O(\rho\,v),\\
 		T_{ij}=&O(\rho\,v^2),
 	\end{split}
 \end{align}
 where  $\rho=\sum\limits_a m_a \delta^{(3)}({\bf x-x}_a)$ and $v$ is the typical velocity of point particles. Hence, the solution for the tensor field to order $O(v^2)$ is
 \begin{align}
 	\begin{split}\label{tensor field}
 		\theta_{00}&=\frac{4}{\phi_0}\sum_a \frac{m_a}{r_a},\\
 		\theta_{0i}&=0,\\
 		\theta_{ij}&=0,
 	\end{split}
 \end{align}
 where $r_{a}=|{\bf x-x}_a|$.\par
 
 $\bullet$ Concerning the scalar field $h_f$,
 to the lowest order, the scalar equation \eqref{Psi} reduces to
 \begin{eqnarray}
 	(\nabla^2 -m_{f}^2)\,h_{f}=\,\frac{8\pi (T\omega_{0}+\phi\,\frac{dT}{d\phi})}{3\phi_{0}\,\omega_{0}},\label{Psii}
 \end{eqnarray}
 where $T$ is given by Eq. \eqref{traceT}.
 To evaluate the right hand side of the above equation, we need to expand the mass of the point particle around the scalar background $\phi_0$
 \begin{equation}\label{massphi}
 	m_a(\phi)=m_a\left[1+s_a\frac{\varphi}{\phi_0}+\frac12(s_a^2+s_a'-s_a)\big(\frac{\varphi}{\phi_0}\big)^2+\cdots\right],
 \end{equation}
 where $m_a\equiv m_a(\phi_0)$ and we have defined $s_a$ and  $s_a'$, the sensitivity and its derivative of point particle $a$, to be \cite{Eardley},
 \begin{equation}
 	s_a=\frac{d \ln m_a(\phi)}{d \ln \phi}\Big|_{\phi_0}, \qquad s_a'=\frac{d^2 \ln m_a(\phi)}{d(\ln \phi)^2}\Big|_{\phi_0} .
 \end{equation}
 Using Eqs. ($\ref{traceT}$) and ($\ref{massphi}$) and expanding the right hand side of Eq. ($\ref{Psii}$) to lowest order, we obtain
 \begin{eqnarray}
 	T\omega_{0}+\phi\,\frac{dT}{d\phi} &=& -\sum_a m_a(\omega_{0}+s_a)\delta^3(\mathbf{x}-\mathbf{x}_a) \label{source1} \\
 	&=& \sum_a m_a(\omega_{0}+s_a)\left(\nabla^2-m_f^2\right)\frac{e^{-m_fr_a}}{4\,\pi\,r_a}.\label{source}
 \end{eqnarray}
 Substituting the above obtained result into Eq. ($\ref{Psii}$), the solution for  $h_f$ will be obtained
 \begin{align}
 	h_{f}=\frac{2}{3\,\omega_{0}\,\phi_{0}\,}\sum_a\,m_a\frac{e^{-m_fr_a}}{r_a}\left(\omega_{0}+s_a\right).
 \end{align}
 $\bullet$ Finally, the equation for the scalar field $\varphi$ to Newtonian order reduces to
 \begin{align}
 	\nabla^2\varphi=\,\frac{8\pi\,T^*}{2\,\omega_{0}+3\,}\,-\,\frac{3\,\phi_0\,}{2\,\omega_{0}+3\,}\, \nabla^2\,h_f.
 \end{align}
 where
 \begin{eqnarray}
 	T^{\star} &= &\sum_a m_a(2s_a-1)\delta^3(\mathbf{x}-\mathbf{x}_a) \label{Tstar1} \\
 	&= &-\frac{1}{4\pi}\sum_a m_a(2s_a-1)\,\nabla^2\,(\frac{1}{r_a}).\label{Tstar2}
 \end{eqnarray}
 and therefore, the solution will be as follows
 \begin{align}
 	\frac{\varphi}{\phi_0}=\frac{-2}{\phi_0\,(2\,\omega_{0}+3\,)}\sum_a\frac{m_a}{r_a}(2\,s_a-1\,+\,\frac{\omega_{0}+s_a}{\omega_{0}}\,e^{-m_fr_a})      .
 \end{align}
 After using the definition of $\theta_{\mu\nu}$ (Eq. ($\ref{new tensor}$)) to find the metric perturbation to Newtonian order and substituting the results into Eq. ($\ref{weak metric}$), the metric coefficients are obtained as follows
 \begin{align}
 	\begin{split}\label{metric}
 		g_{00}&=-1+\frac{1}{\phi_0}\sum_a\frac{2\,m_a}{r_a}\,\left(1\,+\,\frac{1-2\,s_a}{2\,\omega_{0}+3\,}\,+\,\frac{2(\omega_{0}+s_a)}{3\,\,(2\,\omega_{0}+3\,)}e^{-m_fr_a}\right),\\
 		g_{ij}&=\delta_{ij}\left[1+\frac{1}{\phi_0}\sum_a\frac{2\,m_a}{r_a}\,\left(1\,-\,\frac{1-2\,s_a}{2\,\omega_{0}+3\,}\,-\,\frac{2(\omega_{0}+s_a)}{3\,(2\,\omega_{0}+3\,)}e^{-m_fr_a}\right)\right],\\
 		g_{0i}&=0,
 	\end{split}
 \end{align}
 where in the limit $m_{f} \to \infty$, they reduce to those obtained in the massless Brans-Dicke case \cite{califord}.
 
 \subsection{{Equation of motion of point particle}} \label{EOM point mass}
 Using the results obtained in the previous section, one can find the
 EIH equations of motion to describe the approximate dynamics of system \cite{Will:2018bme}. According to the EIH approach, it should be first obtained matter Lagrangian for the $a^{th}$ particle and then made it symmetric under the interchange of all pairs of particles in the system.
 
 From Eq. \eqref{mass action}, the matter Lagrangian for the $a^{th}$ particle in the system is
 given by
 \begin{align}
 	L_a = m_a(\phi)\left(-g_{00}-2g_{0i}v_a^i-g_{ij}v_a^iv_a^j\right)^\frac{1}{2}\,.\label{matter lagrangian}
 \end{align}
 Substituting Eqs. \eqref{metric} into the above relation and making the gravitational terms in $L_a$ symmetric under the interchange of all pairs
 of particles, the matter Lagrangian to Newtonian order reduces to
 \begin{align}
 	L_\mathrm{EIH}=-\sum_a m_a(1-\frac{1}{2}v_a^2)+\frac{1}{2}\sum_{a\neq b}\frac{m_a m_b}{r_{ab}}\mathfrak{g}_{ab},
 \end{align}
 where
 \begin{align}
 	\mathfrak{g}_{ab}=\frac{1}{\phi_0\,}\Bigg[1+\frac{3\,(1-2s_a)(1-2s_b)+2\,e^{-m_fr_{ab}}\,(\omega_{0}+s_a)(\omega_{0}+s_b)}{3(2\omega_{0}+3)}\Bigg]\, .
 \end{align}
 It should be mentioned that the lower indexes $a$ and $b$ refer to the $a^{th}$ and $b^{th}$ particles and they are not tensorial indexes. Employing Euler-Lagrange equations,
 \begin{equation}
 	\frac{d}{dt}(\frac{\partial L}{\partial v_{a}^{i}})-\frac{\partial L}{\partial x_{a}^{i}}=0,
 \end{equation}
 n-particle equations of motion are found to be
 \begin{align}
 	\mathbf{a}_a&=-\sum_{b\neq a}\frac{m_b}{r_{ab}^2}\tilde{\mathfrak{g}}_{ab}\hat{\mathbf{r}}_{ab},
 \end{align}
 where
 \begin{align}
 	\tilde{\mathfrak{g}}_{ab}=\frac{1}{\phi_0\,}\Bigg[1+\frac{3\,(1-2s_a)(1-2s_b)+2(\omega_{0}+s_a)(\omega_{0}+s_b)\,(1+r\,m_f)e^{-m_fr_{ab}}}{3(2\omega_{0}+3)}\Bigg]\,,
 \end{align}
 Since our final aim is to apply this formalism to binary systems
 containing compact objects, we now restrict our attention to two-body systems with the center of mass at the origin. In this regard, we define
 \begin{align}
 	\label{two-body}
 	\mathbf{r}=\mathbf{r}_2-\mathbf{r}_1,\,\,\,\,\,\,\,\,m=m_1+m_2,\,\,\,\,\,\,\,\,\,\mu=\frac{m_1m_2}{m}\,,
 \end{align}
 and also we consider $\tilde{\mathfrak{g}}_{12}=\mathfrak{g}$. Under the above conditions, the equations of motion in the Newtonian limit of the orbital motion will be obtained as
 \begin{align}
 	\label{EOM}
 	\mathbf{a}= & -\frac{m\mathbf{r}}{r^3}\tilde{\mathfrak{g}}\,,
 \end{align}
 where
 \begin{align}\label{gtilde}
 	\tilde{\mathfrak{g}}=\frac{1}{\phi_0\,}\delta\,,
 \end{align}
 and $\delta$ is defined as
 \begin{align}
 	\delta=1+\frac{1}{3(2\omega_{0}+3)}\Big[3\,(1-2s_1)(1-2s_2)+2(\omega_{0}+s_1)(\omega_{0}+s_2)\,(1+r\,m_f)e^{-m_fr_{ab}}\Big].
 \end{align}
 Making use of the above results, we will calculate the gravitational radiation produced by compact binaries in GBD theory in the next section.
 
 
 \section{{Gravitational radiation generated by compact binaries}}\label{GW radiation}
 
 This section is devoted to calculating the gravitational waveforms emitted by a compact binary system and the gravitational radiation power of all propagating degrees of freedom. Since it is obtained from the gravitational wave stress-energy (pseudo) tensor (GW SET), we first try to find a suitable effective energy momentum tensor for GWs in GBD theory.
 
 \subsection{{Energy momentum tensor and the GW radiation power in GBD theory}} \label{Energy momentum}
 
 Although GWs are expected to carry energy and momentum, it is difficult to define a proper energy-momentum tensor for a gravitational field
 due to a consequence of the equivalence principle \cite{Hobson:2006se,Misner:1973prb}.
 However, many efforts have been put forward into finding the proper GW SET and different methods have been proposed to calculate it \cite{Saffer:2017ywl}. Here, we follow the approach of Isaacson which is called {\it{perturbed field equations method}} \cite{Isaacson:1968hbi}. This method is based on perturbing the field equations to second order by considering all propagating degrees of freedom around a generic background.\par
 To this end, one can decompose the metric $g_{\mu\nu}$ in orders of $h_{\mu\nu}$ as
 \begin{equation}
 	g_{\mu\nu} = \eta_{\mu\nu} + {h^{(1)}}_{\mu\nu} + {h^{(2)}}_{\mu\nu} + \ldots,
 	\label{eq:G_exp}
 \end{equation}
 where $\eta_{\mu\nu}$ is the Minkowski metric and ${h^{(1)}}_{\mu\nu}$ and ${h^{(2)}}_{\mu\nu}$ are first order and second order metric perturbations, respectively. Moreover, we need to expand the two scalar fields as follows
 \begin{eqnarray}
 	\phi &=& \phi_{0} + \varphi^{(1)} + \varphi^{(2)},\\
 	f'(R)& = &1 + a_2 R^{(1)}\,+\,\frac{1}{2}\,a_3\,R^{(2)}.
 \end{eqnarray}
 With these decompositions, we now expand the tensor
 field equations $\mathcal{G}_{\mu\nu}=0$ to the second order of
 perturbation, leading to the following relation
 \begin{eqnarray}
 	{\mathcal{G}^{(2)}}_{\mu\nu} && = \phi_{0}\Bigg[ {R^{(2)}}_{\mu\nu} + f'^{(1)}{R^{(1)}}_{\mu\nu}+\frac{\varphi}{\phi_0}{R^{(1)}}_{\mu\nu}-\frac{1}{2}{h^{(1)}}_{\mu\nu}\,f^{(1)} -\frac{1}{2}\eta_{\mu\nu}\,f^{(2)}-\frac{\varphi}{2\phi_0}f^{(1)}\,\eta_{\mu\nu}- \partial_\mu\partial_\nu f'^{(2)}\nonumber\\
 	&& + {{\Gamma^{(1)}}^\lambda}_{\mu\nu}\partial_\lambda f'^{(1)} + \eta_{\mu\nu}\eta^{\alpha\beta}\partial_\alpha\partial_\beta f'^{(2)}+h_{\mu\nu}\eta^{\alpha\beta}\partial_\alpha\partial_\beta f'^{(1)}- {\;} \eta_{\mu\nu}h^{\alpha\beta}\partial_\alpha\partial_\beta f'^{(1)}\nonumber\\
 	&&-\eta_{\mu\nu}\eta^{\alpha\beta}{{\Gamma^{(1)}}^\lambda}_{\alpha\beta}\partial_\lambda f'^{(1)} - \frac{1}{\phi_0}\Big(\partial_\mu\partial_\nu \varphi^{(2)}+\partial_\mu\partial_\nu (\varphi^{(1)}\,f'^{(1)})-{{\Gamma^{(1)}}^\lambda}_{\mu\nu}\partial_\lambda \varphi^{(1)}-\eta_{\mu\nu}\eta^{\alpha\beta}\partial_\alpha\partial_\beta \varphi^{(2)}\nonumber\\
 	&&-h_{\mu\nu}\eta^{\alpha\beta}\partial_\alpha\partial_\beta \varphi^{(1)}+ \eta_{\mu\nu}h^{\alpha\beta}\partial_\alpha\partial_\beta \varphi^{(1)}+\eta_{\mu\nu}\eta^{\alpha\beta}{{\Gamma^{(1)}}^\lambda}_{\alpha\beta}\partial_\lambda \varphi^{(1)}          \Big)\nonumber\\
 	&&-\frac{\omega_{0}}{\phi_{0}^{2}}\Big(\partial_\mu\varphi\partial_\nu\varphi-\frac{1}{2}\eta_{\mu\nu}\,\partial_\alpha\varphi\partial^\alpha\varphi\Big) \Bigg],
 \end{eqnarray}
 where $\mathcal{G}_{\mu\nu}$ is given by Eq. \eqref{gravitational equation}. To find the explicit relation for ${\mathcal{G}^{(2)}}_{\mu\nu}$, we need to calculate the Christoffel symbols as well as the Ricci scalar and Ricci tensor to the second order of perturbations.\par
 Using Eq. \eqref{new tensor} and working on the transverse-traceless (TT) gauge ($\partial^{\mu}\theta_{\mu\nu}=0\,\,\,\&\,\,\theta=\eta_{\mu\nu}\theta^{\mu\nu}=0 $),
 \begin{equation}
 	h_{\mu\nu}=\theta_{\mu\nu}-\eta_{\mu\nu}\frac{\varphi}{\phi_{0}}\,\,-\,\eta_{\mu\nu}h_{f},
 \end{equation}
 the following results will be obtained for the Christoffel symbols
 \begin{eqnarray}
 	{{\Gamma^{(1)}}^\rho}_{\mu\nu} & = & \frac{1}{2}\eta^{\rho\lambda}\left[\partial_\mu \left(\theta_{\lambda\nu} - a_2 R^{(1)}\eta_{\lambda\nu}-\frac{\varphi}{\phi_0}\,\eta_{\nu\lambda}\right) \right. \nonumber \\*
 	& & + \left. \partial_\nu \left(\theta_{\lambda\mu} - a_2 R^{(1)}\eta_{\lambda\mu}-\frac{\varphi}{\phi_0}\,\eta_{\mu\lambda}\right) \right. \nonumber \\*
 	& & - \left. \partial_\lambda \left(\theta_{\mu\nu} - a_2 R^{(1)}\eta_{\mu\nu}-\frac{\varphi}{\phi_0}\,\eta_{\mu\nu}\right)\right],
 \end{eqnarray}
 \begin{eqnarray}
 	{{\Gamma^{(2)}}^\rho}_{\mu\nu} & = & -\frac{1}{2}h^{\rho\lambda}(\partial_\mu h_{\lambda\nu} + \partial_\nu h_{\lambda\mu} - \partial_\lambda h_{\mu\nu}) \nonumber \\*
 	& = & -\frac{1}{2}\left(\theta_{\lambda\nu} - a_2 R^{(1)}\eta_{\lambda\nu}-\frac{\varphi}{\phi_0}\,\eta_{\nu\lambda}\right)\left[\partial_\mu \left(\theta_{\lambda\nu} - a_2 R^{(1)}\eta_{\lambda\nu}-\frac{\varphi}{\phi_0}\,\eta_{\nu\lambda}\right) \right. \nonumber \\*
 	& & + \left. \partial_\nu \left(\theta_{\lambda\mu} - a_2 R^{(1)}\eta_{\lambda\mu}-\frac{\varphi}{\phi_0}\,\eta_{\mu\lambda}\right) \right. - \left. \partial_\lambda \left(\theta_{\mu\nu} - a_2 R^{(1)}\eta_{\mu\nu}-\frac{\varphi}{\phi_0}\,\eta_{\mu\nu}\right)\right].
 \end{eqnarray}
 Furthermore, the expansions of the Ricci tensor are
 \begin{equation}\label{ricci tensor1}
 	{R^{(1)}}_{\mu\nu} = a_2\partial_\mu\partial_\nu R^{(1)} + \partial_\mu\partial_\nu (\frac{\varphi}{\phi_0})+\frac{1}{2}\,a_{2}\eta_{\mu\nu}\,\Box_{\eta}R^{(1)}+\frac{1}{2}\,\eta_{\mu\nu}\,\Box_{\eta}(\frac{\varphi}{\phi_0}),
 \end{equation}
 and
 \begin{eqnarray}\label{ricci tensor2}
 	{R^{(2)}}_{\mu\nu} & = & \partial_\rho {{\Gamma^{(2)}}^\rho}_{\mu\nu} - \partial_\nu {{\Gamma^{(2)}}^\rho}_{\mu\rho} + {{\Gamma^{(1)}}^\rho}_{\mu\nu}{{\Gamma^{(1)}}^\sigma}_{\rho\sigma}
 	- {{\Gamma^{(1)}}^\rho}_{\mu\sigma}{{\Gamma^{(1)}}^\sigma}_{\rho\nu} \nonumber \\
 	& = & \frac{1}{2}\Bigg[\frac{1}{2}\partial_\mu\theta_{\lambda\rho}\partial_\nu\theta^{\lambda\rho} +\theta^{\rho\lambda}\Big(\partial_\mu\partial_\nu\theta_{\rho\lambda}+\partial_\rho\partial_\lambda\theta_{\mu\nu}-\partial_\nu\partial_\rho\theta_{\lambda\mu}-\partial_\rho\partial_\mu\theta_{\lambda\nu}\Big)+\partial_\sigma\theta^{\rho}_{\mu}\Big(\partial^{\sigma}\theta_{\rho\nu}-\partial_{\rho}\theta^{\sigma}_{\nu}\Big)\nonumber\\
 	&& +a_{2}\Big(\theta^{\rho}_{\mu}\partial_{\rho}\partial_{\nu}R^{(1)}-\partial^{\lambda}R^{(1)}\partial_{\lambda}\theta_{\mu\nu}-\eta_{\mu\nu}\theta^{\rho\lambda}\partial_{\rho}\partial_{\lambda}R^{(1)}+\theta^{\rho}_{\nu}\partial_{\rho}\partial_{\mu}R^{(1)}\Big)\nonumber\\
 	&& +a_{2}^{2}\Big(3\partial_{\mu}R^{(1)}\partial_{\nu}R^{(1)}+2R^{(1)}\partial_{\mu}\partial_{\nu}R^{(1)}+\eta_{\mu\nu}R^{(1)}\Box_{\eta}R^{(1)}\Big)\nonumber\\
 	&&+\frac{1}{\phi_{0}}\Big(\theta_{\mu}^{\rho}\partial_{\nu}\partial_{\rho}\varphi+\theta_{\nu}^{\rho}\partial_{\mu}\partial_{\rho}\varphi-\partial_{\lambda}\varphi\partial^{\lambda}\theta_{\mu\nu}-\eta_{\mu\nu}\theta^{\rho\lambda}\partial_{\rho}\partial_{\lambda}\varphi\Big)\nonumber\\
 	&&+\frac{1}{\phi_{0}^{2}}\Big(3\partial_{\mu}\varphi\partial_{\nu}\varphi+2\varphi\partial_{\mu}\partial_{\nu}\varphi+\eta_{\mu\nu}\varphi\Box_{\eta}\varphi\Big) +\frac{a_{2}}{\phi_0}\Big(-2\partial_{\mu}R^{(1)}\partial_{\nu}\varphi-2\partial_{\nu}R^{(1)}\partial_{\mu}\varphi\nonumber\\
 	&&+2\eta_{\mu\nu}\partial_{\lambda}R^{(1)}\partial_{\lambda}\varphi+2R^{(1)}\partial_{\mu}\partial_{\nu}\varphi-2\varphi\partial_{\mu}\partial_{\nu}R^{(1)}+\eta_{\mu\nu}R^{(1)}\Box_{\eta}\varphi+\eta_{\mu\nu}\varphi\Box_{\eta}R^{(1)}\Big)\Bigg],
 \end{eqnarray}
 where we have used the gauge condition \eqref{Lorentz-gauge} to obtain the above results. Besides, Ricci scalar will be found by contracting the Ricci tensor with the full  metric as follows
 \begin{eqnarray}\label{ricci scalar1}
 	R^{(1)}&=&\eta^{\mu\nu} {R^{(1)}}_{\mu\nu} = 3a_{2}\Box_{\eta}R^{(1)}+3\Box_{\eta}(\frac{\varphi}{\phi_0}),
 \end{eqnarray}
 and
 \begin{eqnarray}\label{ricci scalar2}
 	R^{(2)} & = & \eta^{\mu\nu} {R^{(2)}}_{\mu\nu} - h^{\mu\nu} {R^{(1)}}_{\mu\nu} \nonumber \\*
 	& = & \frac{3}{4}\partial_\mu\theta_{\lambda\rho}\partial^\mu\theta^{\lambda\rho} - \frac{1}{2} \partial_{\sigma}\theta^{\rho}_{\mu}\partial_\rho\theta^{\sigma\mu}\nonumber\\
 	&& - 2a_2 \,\theta^{\mu\nu}\,\partial_\mu\partial_\nu R^{(1)}+a_{2}^{2}\Big(\frac{3}{2}\,\partial_{\mu}R^{(1)}\partial^{\mu}R^{(1)}+6\,R^{(1)}\Box_{\eta}R^{(1)}\Big)\nonumber\\
 	&&- \frac{2}{\phi_0}\, \theta^{\mu\nu}\partial_\mu\partial_\nu \varphi\,+\,\frac{1}{\phi_0^2}\,\Big(\frac{3}{2}\,\partial_{\mu}\varphi\,\partial^{\mu}\varphi\,+\,6\varphi\,\Box_{\eta}\varphi\Big)\nonumber\\
 	&& +\frac{a_{2}}{\phi_{0}}\,\Big(2\,\partial_{\mu}R^{(1)}\partial^{\mu}\varphi\,+3\,R^{(1)}\,\Box_{\eta}\varphi\,+\,3\,\varphi\,\Box_{\eta}R^{(1)}\,\Big).
 \end{eqnarray}
 By Combining all obtained results with taking the following relations into account
 \begin{eqnarray}
 	f^{(1)} & = & R^{(1)}, \\
 	f^{(2)} & = & R^{(2)} + \frac{a_2}{2}{R^{(1)}}^2,
 \end{eqnarray}
 \begin{eqnarray}
 	f'^{(0)}&=& 1, \\
 	f'^{(1)} & = & a_2 R^{(1)}, \\
 	f'^{(2)} & = & a_2 R^{(2)} + \frac{a_3}{2}{R^{(1)}}^2,
 \end{eqnarray}
 one can find the explicit relation for $\mathcal{G}_{\mu\nu}$ as
 \begin{eqnarray}\label{seccond metric}
 	\mathcal{G}_{\mu\nu}^{(2)}&=&\phi_{0}\Bigg[R_{\mu\nu}^{(2)}\,+\,\frac{\varphi}{\phi_0}\,R_{\mu\nu}^{(1)}\,-\,\frac{1}{2}\,R^{(1)}\,\theta_{\mu\nu}\,-\,\frac{1}{2}\eta_{\mu\nu}\,R^{(2)}-a_{3}\Big(\frac{1}{2}\partial_{\mu}\partial_{\nu}{R^{(1)}}^2-\frac{1}{2}\eta_{\mu\nu}\Box_{\eta}{R^{(1)}}^2\Big)\nonumber\\
 	&&+a_{2}\Big(R^{(1)}\,R_{\mu\nu}^{(1)}\,+\,\frac{1}{4}\,\eta_{\mu\nu}\,{R^{(1)}}^{2}+\frac{1}{2}\partial^{\rho}R^{(1)}\partial_{\mu}\theta_{\nu\rho}+\frac{1}{2}\partial^{\rho}R^{(1)}\partial_{\nu}\theta_{\mu\rho}-\frac{1}{2}\partial^{\rho}R^{(1)}\partial_{\rho}\theta^{\mu\nu}\nonumber\\
 	&&+\eta_{\mu\nu}\Box_{\eta}R^{(2)}+\theta_{\mu\nu}\Box_{\eta}R^{(1)}-\eta_{\mu\nu}\theta^{\alpha\beta}\partial_{\alpha}\partial_{\beta}R^{(1)}-\eta_{\mu\nu}\partial^{\rho}R^{(1)}\partial^{\alpha}\theta_{\alpha\rho}+\eta_{\mu\nu}\eta^{\alpha\beta}\partial^{\rho}R^{(1)}\partial_{\rho}\theta_{\alpha\beta}\Big)\nonumber\\
 	&&-a_{2}^{2}\Big(\partial_{\mu}R^{(1)}\partial_{\nu}R^{(1)}+\frac{1}{2}\eta_{\mu\nu}\partial^{\rho}R^{(1)}\partial_{\rho}R^{(1)}\Big)-\frac{1}{\phi_{0}^2}\Big(\partial_{\mu}\varphi\partial_{\nu}\varphi+\frac{1}{2}\eta_{\mu\nu}\partial_{\rho}\varphi\partial^{\rho}\varphi\Big)\nonumber\\
 	&&-\frac{1}{\phi_0}\Big(\partial_{\mu}\partial_{\nu}\varphi^{(2)}-\frac{1}{2}\partial^{\rho}\varphi\partial_{\mu}\theta_{\nu\rho}-\frac{1}{2}\partial^{\rho}\varphi\partial_{\nu}\theta_{\mu\rho}+\frac{1}{2}\partial^{\rho}\varphi\partial_{\rho}\theta_{\mu\nu}-\eta_{\mu\nu}\Box_{\eta}\varphi^{(2)}-\theta_{\mu\nu}\Box_{\eta}\varphi\nonumber\\
 	&&+\eta_{\mu\nu}\theta^{\alpha\beta}\partial_{\alpha}\partial_{\beta}\varphi+\eta_{\mu\nu}\partial^{\rho}\varphi\partial^{\alpha}\theta_{\alpha\rho}-\frac{1}{2}\eta_{\mu\nu}\eta^{\alpha\beta}\partial^{\rho}\varphi\partial_{\rho}\theta_{\alpha\beta}\Big)\nonumber\\
 	&&-\frac{a_{2}}{\phi_0}\Big(\partial_{\mu}\phi\partial_{\nu}R^{(1)}+\partial_{\nu}\varphi\partial_{\mu}R^{(1)}+\frac{1}{2}\eta_{\mu\nu}\partial_{\rho}R^{(1)}\partial^{\rho}\varphi+\partial_{\mu}\partial_{\nu}(\varphi R^{(1)})\Big)\nonumber\\
 	&&-\frac{\omega_{0}}{\phi_{0}^2}\Big(\partial_{\mu}\varphi\partial_{\nu}\varphi-\frac{1}{2}\eta_{\mu\nu}\partial^{\rho}\varphi\partial_{\rho}\varphi\Big)\Bigg].
 \end{eqnarray}
 The effective GW SET for the radiation will be obtained by averaging the above relation over several wavelengths \cite{Misner:1973prb, Saffer:2017ywl}
 \begin{equation}
 	t_{\mu\nu} = -\frac{1}{8\pi }\left\langle{\mathcal{G}^{(\text{2})}}_{\mu\nu}\right\rangle,
 \end{equation}
 where the angled-bracket stands for short wavelength averaging.
 Due to the averaging over all directions at each point,
 the averages of the first derivatives vanish as $ \left\langle\partial_\mu V\right\rangle = 0$, which also implies that $\left\langle U\partial_\mu V\right\rangle = -\left\langle V \partial_\mu U\right\rangle $ \cite{Misner:1973prb}. Substituting Eqs.\eqref{ricci tensor1}, \eqref{ricci tensor2}, \eqref{ricci scalar1} and \eqref{ricci scalar2} into Eq.\eqref{seccond metric}
 and repeated application of the mentioned point along with using gauge condition \eqref{Lorentz-gauge} and wave equations \eqref{theta}, \eqref{phi} and \eqref{Psi} yield
 \begin{eqnarray}
 	\left\langle {\mathcal{G}^{(2)}}_{\mu\nu}\right\rangle &=& \phi_{0}\Bigg[\left\langle -\frac{1}{4} \partial_\mu\theta_{\lambda\rho}^{TT}\partial_\nu\theta^{\lambda\rho}_{TT} \right\rangle- a_{2}^{2}\left\langle\frac{3}{2}\partial_\mu R^{(1)}\partial_\nu R^{(1)}+\frac{27}{4(2\omega_{0}+3)}\eta_{\mu\nu}R^{(1)}\Box_{\eta}R^{(1)}\right\rangle\nonumber\\
 	&&-\frac{2\omega_{0}+3}{2\phi_{0}^2}\left\langle\partial_{\mu}\varphi\partial_{\nu}\varphi+\frac{\eta_{\mu\nu}}{2}\varphi\Box_{\eta}\varphi\right\rangle -\frac{a_{2}}{\phi_{0}}\left\langle 8\partial_{\mu}R^{(1)}\partial_{\nu}\varphi-\frac{\eta_{\mu\nu}}{2}R^{(1)}\Box_{\eta}\varphi\right\rangle\Bigg],
 \end{eqnarray}
 where $\theta_{\lambda\rho}^{TT}$ recall that the tensor field $\theta_{\lambda\rho}$ must be computed in the TT gauge.
 Hence, the effective GW SET for the radiation reads
 \begin{eqnarray}\label{GW SET}
 	t_{\mu\nu}  &=& \frac{\phi_{0}}{32\pi\,}\Big[\left\langle \partial_\mu\theta_{\lambda\rho}^{TT}\partial_\nu\theta^{\lambda\rho}_{TT} \right\rangle+ a_{2}^{2}\left\langle\,6\,\partial_\mu R^{(1)}\partial_\nu R^{(1)}+\frac{27}{(2\omega_{0}+3)}\eta_{\mu\nu}R^{(1)}\Box_{\eta}R^{(1)}\right\rangle\nonumber\\
 	&&+\frac{4\omega_{0}+6}{\phi_{0}^2}\left\langle\partial_{\mu}\varphi\partial_{\nu}\varphi+\frac{\eta_{\mu\nu}}{2}\varphi\Box_{\eta}\varphi\right\rangle +\frac{4 a_{2}}{\phi_{0}}\left\langle\partial_{\mu}R^{(1)}\partial_{\nu}\varphi-\frac{\eta_{\mu\nu}}{2}R^{(1)}\Box_{\eta}\varphi\right\rangle\Big]
 \end{eqnarray}
 Now, we are in a position to calculate the power radiated of GWs . Considering that the GW radiation power is given by \cite{Maggiore:2007ulw}
 \begin{equation}
 	\frac{dE_{\mathrm{GW}}}{dt}=\int d \Omega\, \mathfrak{R}^2\, t^{0R}
 \end{equation}
 and taking Eq. \eqref{GW SET} into account, the GW radiation power in GBT theory can be calculated as
 \begin{eqnarray}\label{Edot}
 	\dot{E}_{\mathrm{GW}}&=&\int d \Omega\,         \frac{\phi_{0}\,\mathfrak{R}^{2}}{32\pi\,}\Big[\left\langle -{\theta^{TT}_{ij}}_{,0}\,{\theta^{TT}_{ij}}_{,0} \right\rangle+ 6\,a_{2}^{2}\left\langle\, R^{(1)}_{,0} R^{(1)}_{,\mathfrak{R}}\right\rangle-\frac{4\omega_{0} +6}{\phi_{0}^2}\left\langle\varphi_{,0}\varphi_{,0}\right\rangle +\frac{4\,a_{2}}{\phi_{0}}\left\langle\,R^{(1)}_{,0}{\varphi}_{,\mathfrak{R}}\right\rangle\Big]\nonumber\\
 \end{eqnarray}
 where $d\Omega$ indicates the solid angle element and $\mathfrak{R}$ is the coordinate distance of the field point relative to the center of mass of the source. Also, note that to arrive the above relation, we have used the fact that for the tensor gravitational waves and the scalar field that are massless  $t^{0\mathfrak{R}}=t^{00}+O(\frac{1}{\mathfrak{R}^3})$ \cite{Maggiore:2007ulw} while for the massive scalar field,  $t^{0\mathfrak{R}}\neq t^{00}+O(\frac{1}{\mathfrak{R}^3})$.\par
 According to the above results, we will compute the power radiated in GWs due to the tensor field as well as both scalar fields radiation.
 
 \subsection{{ Calculation of the GW radiation power due to the tensor field}}\label{GW radiation1}
 
 In order to find the tensor field contribution to the GW radiation power of a binary system, we need to obtain the solution of the tensor wave equation \eqref{theta}. To this end, one can use the method of Green's function and multipole expansion, and hence the tensor wave equation (to the leading order) takes the form
 \begin{align}
 	\label{thetaij}
 	\theta^{ij} =\frac{2}{\mathfrak{R}}\frac{\partial^2}{\partial t^2}\int\tau^{00}(t-\mathfrak{R},\mathbf{r'})r'^{i}r'^{j}d^3\mathbf{r'},
 \end{align}
 where
 \begin{align}
 	\tau^{00}=\frac{1}{\phi_0\,}\sum_am_a\delta^3(\mathbf{r'-r_a})\,.
 \end{align}
 Specializing in a two body system in the center of the mass frame, we have
 \begin{align}
 	\label{theta_harmonic}
 	\theta^{ij}(t,\mathbf{\mathfrak{R}})=\frac{4}{\phi_0\,}\mathfrak{R}^{-1}\mu\Big(v^iv^j-\tilde{\mathfrak{g}} m\frac{r^i r^j}{r^3}\Big)\,,
 \end{align}
 where  $v^i\equiv v_1^i-v_2^i$ is the relative velocity and
 we have used Eq.\eqref{EOM} to substitute $\ddot{r}^i$. To project Eq. \eqref{theta_harmonic} onto the TT gauge, one can use the following projection operator \cite{Hobson:2006se, Maggiore:2007ulw}
 \begin{eqnarray}
 	\Lambda(\mathbf{\hat{n}})_{ij,kl}&=& \delta_{ik}\delta_{jl}-\frac{1}{2}\delta_{ij}\delta_{kl}-n_in_k\delta_{jl}-n_jn_l\delta_{ik} \nonumber \\
 	&+&\frac{1}{2}n_kn_l\delta_{ij}+\frac{1}{2}n_in_j\delta_{kl}+\frac{1}{2}n_in_jn_kn_l\,,
 \end{eqnarray}
 so that
 \begin{equation}
 	\theta^{ij}_{TT}=\Lambda(\mathbf{\hat{n}})_{ij,kl}\theta^{kl}\,.
 \end{equation}
 According to Eq.\eqref{Edot}, the GW radiation power due to the tensor field is given by
 \begin{equation}
 	\dot{E}^{t}_{\mathrm{GW}}=\,-\,\int d\Omega \,\frac{\phi_0\,\mathfrak{R}^{2}}{32\pi\,}\langle{\theta^{TT}_{ij}}_{,0}{\theta^{TT}_{ij}}_{,0}\rangle\,=\,-\frac{\mathfrak{R}^2}{32\pi}\phi_0\Big\langle
 	\int\,d\Omega\,\Lambda_{kl,mn}\theta^{kl}_{,0}\theta^{mn}_{,0}\Big\rangle,
 \end{equation}
 where we have used the identity $\Lambda_{ij,kl}\Lambda_{kl,nm}=\Lambda_{ij,nm}$ \cite{Maggiore:2007ulw} in the second equality. Since the only angular dependence in the integrand of the above relation is contained in $\Lambda_{ij,mn}$, one can use the following identity \cite{Maggiore:2007ulw}
 \begin{equation}
 	\int\Lambda_{ij,mn}d\Omega=\frac{2\pi}{15}\big(11\delta_{im}\delta_{jn}-4\delta_{ij}\delta_{mn}+\delta_{in}\delta_{jm}\big)\,,
 \end{equation}
 and the result of the contribution of the tensor field in the GW radiation power yields
 \begin{equation}\label{Edott}
 	\dot{E}^{t}_{\mathrm{GW}}=-\frac{\mathfrak{R}^2}{60\,}\phi_0\Big\langle3\;\theta^{ij}_{\;\;,0}\theta^{ij}_{\;\;,0}-\;\theta^k_{\;k,0}\theta^k_{\;k,0}\Big\rangle\,.
 \end{equation}
 To obtain the final result for the power emitted in tensor GWs, we need to perform the average over one period. For this purpose, we will consider a special case in which the orbital motion is circular and can be parameterized as follows
 \begin{align}
 	\label{circle}
 	& r_1=r\;\mathrm{cos}\big(\upvarpi (t-\mathfrak{R})\big),\,\,\,\,\,\,\,\,\,\,\,\,\,\,\,\, r_2=r\;\mathrm{sin}\big(\upvarpi (t-\mathfrak{R})\big),\,\,\,\,\,\,\,\,\,\,\,\,\,\,\,\,\,\,r_3=0\,, \nonumber \\
 	& v_1=-v\;\mathrm{sin}\big(\upvarpi (t-\mathfrak{R})\big),\,\,\,\,\,\,\,\,\,\,\,\,\,\, v_2=v\;\mathrm{cos}\big(\upvarpi (t-\mathfrak{R})\big),\,\,\,\,\,\,\,\,\,\,\,\,\,\,\,\, v_3=0,\,
 \end{align}
 where $\upvarpi$ is the orbital frequency. Moreover, we suppose
 that the mass of the geometrical scalar field $h_{f}$ is either sufficiently large or
 sufficiently small so that the variations of $\tilde{\mathcal{G}}$ over an
 orbital period can be neglected. With the mentioned two approximations, Eq.\eqref{Edott} reduces to
 \begin{equation}
 	\dot{E}^{t}_{\mathrm{GW}}=-\frac{32}{15}\,\frac{1}{\phi_{0}}\frac{\tilde{\mathfrak{g}}^2\mu^2m^2v^2}{r^4}\,.
 \end{equation}
 Before going further, it will be convenient to define some auxiliary combinations that contain $\omega_{0}$:
 \begin{equation}\label{paras}
 	G\equiv\frac{1}{\phi_0}\frac{4+2\omega_0}{3+2\omega_0},\qquad \xi\equiv\frac{1}{2\omega_0+4},\qquad G(1-\xi)=\frac{1}{\phi_0}.
 \end{equation}
 Taking into account the above relations as well as Eq. \eqref{EOM} and making use of $v=(\tilde{\mathfrak{g}}m\,\upvarpi)^{1/3}$, which helps us to eliminate $v$ in favor of $\upvarpi$, we obtain the final result for
 the power emitted in tensor GWs as
 \begin{equation}\label{Etensor}
 	\dot{E}^{t}_{\mathrm{GW}}=\,-\,\frac{32}{5}\phi_{0}\tilde{\delta}^2(G\,M_c\,\upvarpi)^{10/3},
 \end{equation}
 where $M_c = \mu^{3/5}m^{2/5}$ stands for the chirp mass of the binary system and $\tilde{\delta}$ is defined as $\tilde{\delta}=(1-\xi)^{5/3}\delta^{2/3}$.\par
 Before concluding this subsection, it should be emphasized that the validity of the above result is limited to the cases in which $m_f$ is such that either $e^{-m_fr}\approx 1$ or $e^{-m_fr}\rightarrow 0$.
 \subsection{Calculation of the GW radiation power due to scalar fields}\label{GW radiation2}
 Now, we try to find the scalar field contributions to the GW radiation power of a binary system. To this end, one first needs to get the solutions of the wave equations \eqref{phi} and \eqref{Psi}, which will be calculated in the following:
 
 \subsubsection{Calculation of the contribution due to the geometrical scalar field in the GW radiation power}
 
 According to Eq.\eqref{Edot}, the GW radiation power due to the geometrical scalar field is given by
 \begin{eqnarray}\label{EdotR}
 	\dot{E}_{\mathrm{GW}}^{\Phi}&=&\frac{6\,a_{2}^{2}\phi_{0}}{32\pi\,} \int d \Omega\,\mathfrak{R}^{2}\left\langle\, R^{(1)}_{,0} R^{(1)}_{,\mathfrak{R}}\right\rangle,
 \end{eqnarray}
 Therefore, we first need to obtain the solution of the geometrical scalar wave equation. Making use of the fact that $h_{f}\equiv\delta\Phi^{(1)} = a_{2}R^{(1)}$ as well as $m_{f}^2\equiv\frac{{R}^{\left(1\right)}}{3\,\delta \Phi^{(1)}}\,\frac{2\omega_{0}+3}{2\omega_{0}}$, Eq. \eqref{Psi} reduces to the following equation
 \begin{eqnarray}
 	\Box_{\eta} R^{(1)}-m_{f}^2\,R^{(1)}=\,-16\,\pi\,S,\label{Psi1}
 \end{eqnarray}
 where $m_{f}^{2}=\frac{2\omega_{0}+3}{6\omega_{0} a_{2}}$ and $S= \,\frac{- (T\omega_{0}+\phi\,\frac{dT}{d\phi})}{6\,a_{2}\phi_{0}\,\omega_{0}}$. To solve the above equation, we utilize the retarded Green's function for the massive wave operator
 \begin{equation}
 	(\square_\eta -m_f^2)\mathcal{G}(x)=-4\pi \delta^{(4)}(x)
 \end{equation}
 in which $\delta^{(4)}(x)$ is the four dimensional delta function and the retarded Green's function reads as \cite{Poisson:2011nh}
 \begin{eqnarray}
 	\mathcal{G}(t-t',\mathbf{\mathfrak{R}-r'})= \frac{\delta(t-t'-|\mathbf{\mathfrak{R}-r'}|)}{|\mathbf{\mathfrak{R}-r'}|}
 	-\Theta(t-t'-|\mathbf{\mathfrak{R}-r'}|)\frac{m_fJ_1(m_f\sqrt{(t-t')^2-|\mathbf{\mathfrak{R}-r'}|^2})}{\sqrt{(t-t')^2-|\mathbf{\mathfrak{R}-r'}|^2})},\nonumber\\
 \end{eqnarray}
 where $\Theta$ stands for the Heaviside function and $J_1$ denotes the first kind Bessel function of order one.
 Now, the general solution of the scalar wave equation \eqref{Psi1} can be written as
 \begin{equation}\label{R1}
 	R^{(1)}=R^{(1)}_{B}+R^{(1)}_{m},
 \end{equation}
 where
 \begin{eqnarray}
 	\label{geoscalar1}
 	R^{(1)}_B(t,\mathbf{\mathfrak{R}}) &=& 4\int\int_\mathcal{N} \frac{S(t',\mathbf{r'})\delta(t-t'-|\mathbf{\mathfrak{R}-r'}|)}{|\mathbf{\mathfrak{R}-r'}|}d^3\mathbf{r'}dt'\,,
 \end{eqnarray}
 \begin{eqnarray}    \label{geoscalar2}
 	R^{(1)}_m(t,\mathbf{\mathfrak{R}}) &=& -4\int\int_\mathcal{N} \frac{m_f S(t',\mathbf{r'})
 		J_1(m_f\sqrt{(t-t')^2-|\mathbf{\mathfrak{R}-r'}|^2})}{\sqrt{(t-t')^2-|\mathbf{\mathfrak{R}-r'}|^2}}
 	\Theta(t-t'-|\mathbf{\mathfrak{R}-r'}|)d^3\mathbf{r'}dt' \nonumber \\
 	&=& -4\int_\mathcal{N} d^3\mathbf{r'}
 	\int_{0}^\infty\frac{J_1(z)S(t-\sqrt{|\mathbf{\mathfrak{R}-r'}|^2+(\frac{z}{m_f})^2},\mathbf{r'})}{\sqrt{|\mathbf{\mathfrak{R}-r'}|^2+(\frac{z}{m_f})^2}}dz\,,
 \end{eqnarray}
 where the spatial integration is taken over the near zone $\mathcal{N}$. Besides, we have made the substitution
 $z=m_f\sqrt{(t-t')^2-|\mathbf{\mathfrak{R}-r'}|^2}$ in
 the last line. Considering the field point being in the radiation zone
 ($|\mathbf{\mathfrak{R}}|\gg|\mathbf{r'}|$) and keeping only the leading order $O(\frac{1}{\mathfrak{R}})$ part, we can expand the $\mathbf{r'}$ dependence of the
 integrand and obtain the multipole expansion of the general scalar wave solutions as
 \begin{eqnarray}
 	\label{weak_scalar_exp_sol_massless}
 	R^{(1)}_B&=&\frac{4}{\mathfrak{R}}\sum_{m=0}^\infty\frac{1}{m!}\frac{\partial^m}{\partial t^m}\int_\mathcal{M}d^3\mathbf{r'}S(t-\mathfrak{R},\mathbf{r'})(\mathbf{n \cdot r'})^m\,, \\
 	\label{weak_scalar_exp_sol_massive}
 	R^{(1)}_m&=&
 	-\frac{4}{\mathfrak{R}}\sum_{m=0}^\infty\frac{1}{m!}\frac{\partial^m}{\partial t^m}\int_\mathcal{M}d^3\mathbf{r'}(\mathbf{n\cdot r'})^m \int_0^\infty dz\frac{S(t-\sqrt{\mathfrak{R}^2+(\frac{z}{m_f})^2},\mathbf{r'})J_1(z)}{(1+(\frac{z}{m_f \mathfrak{R}})^2)^\frac{m+1}{2}},
 \end{eqnarray}
 where $\mathbf{n}=\vec{\mathfrak{R}}/\mathfrak{R}$. Substituting Eq. \eqref{source} into the above equations and keeping only the terms up to order $O(\frac{m}{\mathfrak{R}})$ in the scalar wave, one can get the following relations for a binary system
 \begin{equation}\label{Rb}
 	R^{(1)}_B=\frac{2\alpha}{\mathfrak{R}}\Bigg[m\omega_{0}+m_{1}s_{1}+m_{2}s_{2}+\mu\,\mathcal{S}(\mathbf{n}\cdot \mathbf{v})+\mathcal{O}(\frac{m}{r})\Bigg],
 \end{equation}
 \begin{equation}\label{Rm}
 	R^{(1)}_m=-\frac{2\alpha}{\mathfrak{R}}\Bigg[m\omega_{0}+m_{1}s_{1}+m_{2}s_{2}+ \mu\mathcal{S}I_2\big[\mathbf{n}\cdot \mathbf{v}\big]\Bigg],
 \end{equation}
 where
 \begin{eqnarray}
 	\alpha=\frac{1}{3a_{2}\Phi_{0}\omega_{0}},\\
 	\mathcal{S}=s_{1}-s_{2},
 \end{eqnarray}
 and
 the terms $I_n[f(t)]$ denotes the following integrals \cite{Alsing:2011er}
 \begin{equation}
 	I_n[f(t)]\equiv \int_0^\infty dz ~\frac{f(t-\mathfrak{R}u)J_1(z)}{u^n}
 \end{equation}
 with  $u\equiv\sqrt{1+(\frac{z}{m_f \mathfrak{R}})^2}$. Using Eqs. \eqref{R1}, \eqref{Rb} and \eqref{Rm} and taking partial derivative, one finds
 \begin{equation}\label{R0}
 	R^{(1)}_{,0}=-\,\frac{2\alpha \tilde{\mathfrak{g}}m\mu\mathcal{S}}{\mathfrak{R}}\left(\frac{\mathbf{n}\cdot\mathbf{r}}{r^3}-I_2[\frac{\mathbf{n}\cdot\mathbf{r}}{r^3}]\right),
 \end{equation}
 \begin{equation}\label{RR}
 	R^{(1)}_{,\mathfrak{R}}=\,\frac{2\alpha \tilde{\mathfrak{g}}m\mu\mathcal{S}}{\mathfrak{R}}\left(\frac{\mathbf{n}\cdot\mathbf{r}}{r^3}-I_3[\frac{\mathbf{n}\cdot\mathbf{r}}{r^3}]\right).
 \end{equation}
 Making use of the above relations and performing the integration of Eq. \eqref{EdotR} over the solid angle, we get
 \begin{eqnarray}
 	\begin{split}
 		\int d\Omega\langle R^{(1)}_{,0}R^{(1)}_{,\mathfrak{R}}\rangle&=-\left(\frac{2\alpha\tilde{\mathfrak{g}}m\mu}{\mathfrak{R}}\right)^2\int d\Omega\langle \mathcal{S}^2\left(\frac{\mathbf{n}\cdot\mathbf{r}}{r^3}-I_2[\frac{\mathbf{n}\cdot\mathbf{r}}{r^3}]\right)\left(\frac{\mathbf{n}\cdot\mathbf{r}}{r^3}-I_3[\frac{\mathbf{n}\cdot\mathbf{r}}{r^3}]\right)\rangle\nonumber\\
 		&= -\frac{4}{3}\left(\frac{2\alpha\tilde{\mathfrak{g}}m\mu}{\mathfrak{R}}\right)^2\mathcal{S}^2[1-\cos(\upvarpi \mathfrak{R})(C_2(\mathfrak{R};\upvarpi)+C_3(\mathfrak{R};\upvarpi))-\sin(\upvarpi \mathfrak{R})(S_2(\mathfrak{R};\upvarpi)\nonumber\\
 		&+S_3(\mathfrak{R};\upvarpi))+C_2(\mathfrak{R};\upvarpi)C_3(\mathfrak{R};\upvarpi)+S_2(\mathfrak{R};\upvarpi)S_3(\mathfrak{R};\upvarpi)],
 	\end{split}
 \end{eqnarray}
 where
 \begin{equation}
 	C_n(\mathfrak{R};\upvarpi)=\int_0^\infty dz \cos(\upvarpi \mathfrak{R} u)\frac{J_1(z)}{u^n}, \qquad S_n(\mathfrak{R};\upvarpi)=\int_0^\infty dz \sin(\upvarpi \mathfrak{R} u)\frac{J_1(z)}{u^n}.
 \end{equation}
 Since we are interested in the gravitational radiation in
 the far zone, we only need to determine
 the asymptotic expansion of $C_n$ and $S_n$ for $\mathfrak{R}\to \infty$ which will be obtained as follows \cite{Alsing:2011er}
 \begin{equation}\label{identity1}
 	C_n(\mathfrak{R};\upvarpi)\sim\begin{cases}
 		\cos(\upvarpi \mathfrak{R})-\left(\frac{\sqrt{\upvarpi^2-m_f^2}}{\upvarpi}\right)^{n-1}\cos(\mathfrak{R}\sqrt{\upvarpi^2-m_f^2}), & \upvarpi>m_f\\
 		\cos(\upvarpi \mathfrak{R})-\left(\frac{\sqrt{m_f^2-\upvarpi^2}}{\upvarpi}\right)^{n-1}e^{-\mathfrak{R}\sqrt{m_f^2-\upvarpi^2}}\cos\frac{(n-1)\pi}{2},&  \upvarpi<m_f
 	\end{cases}
 \end{equation}
 \begin{equation}\label{identity2}
 	S_n(\mathfrak{R};\upvarpi)\sim\begin{cases}
 		\sin(\upvarpi \mathfrak{R})-\left(\frac{\sqrt{\upvarpi^2-m_f^2}}{\upvarpi}\right)^{n-1}\sin(\mathfrak{R}\sqrt{\upvarpi^2-m_f^2}), & \upvarpi>m_f\\
 		\sin(\upvarpi \mathfrak{R})-\left(\frac{\sqrt{m_f^2-\upvarpi^2}}{\upvarpi}\right)^{n-1}e^{-\mathfrak{R}\sqrt{m_f^2-
 				\upvarpi^2}}\sin\frac{(n-1)\pi}{2},&  \upvarpi<m_f
 	\end{cases}
 \end{equation}
 Making use of these results, we arrive at the following relation for the scalar radiation power
 \begin{equation}\label{des1}
 	\dot{E}_{\mathrm{GW}}^{\Phi}=\frac{8}{9}G(1-\xi)\frac{\tilde{\mathfrak{g}}^2 m^2 \mu^2}{\omega_{0}^{2}r^4}\mathcal{S}^2\left(\frac{\sqrt{\upvarpi^2-m_f^2}}{\upvarpi}\right)^3\Theta(\upvarpi-m_f),
 \end{equation}
 which, by using Eq. \eqref{EOM}, it reduces to
 \begin{equation}\label{des11}
 	\dot{E}_{\mathrm{GW}}^{\Phi}=\frac{8}{9\,\omega_{0}^{2}}\Big(GM_{c}(1-\xi)\Big)^{5/3}\delta^{2/3} \mu\,\mathcal{S}^2\upvarpi^{8/3}\left(\frac{\sqrt{\upvarpi^2-m_f^2}}{\upvarpi}\right)^3\Theta(\upvarpi-m_f).
 \end{equation}
 This result represents the fact that the geometrical scalar field has a dipole contribution to the GW radiation.

 \subsubsection{Calculation of the contribution due to BD scalar field in the GW radiation power}
 
 Making use of the result of the previous section as well as Eqs. \eqref{source1} and \eqref{Tstar1}, the linearized scalar field equation \eqref{phi} reduces to
 \begin{eqnarray}
 	\Box_{\eta} \varphi=\,-16\,\pi\,S',\label{phi1}
 \end{eqnarray}
 where
 \begin{eqnarray}\label{sprime}
 	S'&=&-\frac{2\omega_{0}+1}{2\omega_{0}(2\omega_{0}+3)}\sum_a m_a\,s_a\delta^3(\mathbf{x}-\mathbf{x}_a)\nonumber\\
 	&&+\frac{1}{48\pi \mathfrak{R} a_{2}\omega_{0}^2}\Big[m\omega_{0}+m_{1}s_{1}+m_{2}s_{2}+\mu\,\mathcal{S}(\mathbf{n}\cdot \mathbf{v})-\big(m\omega_{0}+m_{1}s_{1}+m_{2}s_{2}\big)I_{1}[1]-\mu\mathcal{S}I_2[\mathbf{n}\cdot \mathbf{v}]\Big]\nonumber\\
 \end{eqnarray}
 Now, to obtain the solution of the scalar wave equation Eq.\eqref{phi1}, we use the Green's function $\mathcal{G}(t,\mathbf{\mathfrak{R}})= \frac{\delta(t-\mathfrak{R})}{\mathfrak{R}}$ which satisfies
 \begin{equation}
 	\square_\eta \mathcal{G}(x)=-4\pi \delta^{(4)}(x).
 \end{equation}
 By taking the steps mentioned in the previous section, the multipole expansion of the general scalar wave solutions is obtained as
 \begin{eqnarray}
 	\label{weak_scalar_exp_sol_massless1}
 	\varphi&=&\frac{4}{\mathfrak{R}}\sum_{m=0}^\infty\frac{1}{m!}\frac{\partial^m}{\partial t^m}\int_\mathcal{M}d^3\mathbf{r'}S'(t-\mathfrak{R},\mathbf{r'})(\mathbf{n \cdot r'})^m,
 \end{eqnarray}
 in which we consider the field point to be in the radiation zone
 ($|\mathbf{\mathfrak{R}}|\gg|\mathbf{r'}|$) and we have only kept up to the leading order $\mathcal{O}(\frac{1}{\mathfrak{R}})$ part. Since we are interested in finding the scalar wave $\varphi$ to lowest order, the first term of Eq.\eqref{sprime} will be considered. Using Eq. \eqref{weak_scalar_exp_sol_massless1}, the solution of Eq. \eqref{phi1} will be obtained as
 \begin{equation}\label{phi2}
 	\varphi=-\frac{2(2\omega_{0}+1)}{\mathfrak{R}\omega_{0}(2\omega_{0}+3)}\Big[m_{1}s_{1}+m_{2}s_{2}+\mu\,\mathcal{S}(\mathbf{n}\cdot \mathbf{v})\Big],
 \end{equation}
 where it has been specialized for a binary system. Furthermore, taking the partial time derivative of $\varphi$, we find
 \begin{equation}\label{phi0}
 	\varphi_{,0}=\frac{2(2\omega_{0}+1)}{\mathfrak{R}\omega_{0}(2\omega_{0}+3)}\tilde{\mathfrak{g}}m\mu\mathcal{S}\left(\frac{\mathbf{n}\cdot\mathbf{r}}{r^3}\right).
 \end{equation}
 After getting the mentioned results, we are ready to calculate the contribution of the BD scalar field to the GW radiation power. According to Eq.\eqref{Edot}, such a contribution  will be obtained from the following relation
 \begin{eqnarray}\label{Edotphi}
 	\dot{E}_{\mathrm{GW}}^{\varphi}&=&-\frac{2\omega_{0} +3}{16\pi\phi_{0}}\int d \Omega\, \mathfrak{R}^{2}\left\langle\varphi_{,0}\varphi_{,0}\,\right\rangle.
 \end{eqnarray}
 Substituting Eq. \eqref{phi0} into the above relation and performing the integral over the solid angle via the identity
 \begin{align}
 	\label{identity3}
 	\oint n^{i_1}n^{i_2}\cdots n^{i_{2l}}d\Omega=\frac{4\pi\delta^{(i_1i_2}\delta^{i_3i_4}\cdots\delta^{i_{2l-1}i_{2l})}}{(2l+1)!!}\,,
 \end{align}
 we obtain
 \begin{equation}\label{Edotphi1}
 	\dot{E}_{\mathrm{GW}}^{\varphi}=-\frac{(2\omega_{0}+1)^2}{3\,\omega_{0}^{2}(2\omega_{0}+3)}\Big(GM_{c}(1-\xi)\Big)^{5/3}\delta^{2/3} \mu\,\mathcal{S}^2\upvarpi^{8/3},
 \end{equation}
 which shows that the BD scalar field has a dipole contribution to the GW radiation.

 \subsubsection{Calculation of the contribution due to the interaction term between the BD and the geometrical scalar field in the GW radiation power}
 
 In this section, we aim to calculate the energy loss rate due to the coupling term between the BD and the geometrical scalar field, which according to Eq. \eqref{Edot} is obtained as
 \begin{eqnarray}\label{coupling}
 	\dot{E}_{\mathrm{GW}}^{\Phi\varphi}&=&\,\frac{a_{2}}{8\pi}\,\int d \Omega\,  \mathfrak{R}^2   \left\langle\,R^{(1)}_{,0}{\varphi}_{,\mathfrak{R}}\right\rangle.
 \end{eqnarray}
 By considering the fact that for a massless scalar wave $\varphi_{,0}=-\varphi_{,\mathfrak{R}}+O(\frac{1}{\mathfrak{R}^2})$, the above relation reduces to
 \begin{eqnarray}\label{coupling1}
 	\dot{E}_{\mathrm{GW}}^{\Phi\varphi}&=&\,-\frac{a_{2}}{8\pi}\,\int d \Omega\,  \mathfrak{R}^2   \left\langle\,R^{(1)}_{,0}{\varphi}_{,0}\right\rangle
 \end{eqnarray}
 Now, substituting Eqs. \eqref{R0} and \eqref{phi0} into Eq. \eqref{coupling1} yields
 \begin{align}
 	\begin{split}
 		\dot{E}_{\mathrm{GW}}^{\Phi\varphi}=\frac{(2\omega_{0}+1)}{6\phi_{0}\pi\,\omega_{0}^{2}(2\omega_{0}+3)}\big(\mu m \mathcal{S}\tilde{\mathfrak{g}}\big)^2\int d\Omega\langle \left(\frac{\mathbf{n}\cdot\mathbf{r}}{r^3}\right)\left(\frac{\mathbf{n}\cdot\mathbf{r}}{r^3}-I_2[\frac{\mathbf{n}\cdot\mathbf{r}}{r^3}]\right)\rangle.
 	\end{split}
 \end{align}
 To perform the above integral, we use Eqs. \eqref{identity1}, \eqref{identity2} and \eqref{identity3} which lead to obtain the following dipole contribution
 \begin{align}\label{Edotcoup}
 	\begin{split}
 		\dot{E}_{\mathrm{GW}}^{\Phi\varphi}=\frac{2(2\omega_{0}+1)}{9\,\omega_{0}^{2}(2\omega_{0}+3)}\Big(GM_{c}(1-\xi)\Big)^{5/3}\delta^{2/3} \mu\,\mathcal{S}^2\upvarpi^{8/3}\,\cos (\mathfrak{R}\frac{m_{f}^2}{\upvarpi})\left(\frac{\sqrt{\upvarpi^2-m_f^2}}{\upvarpi}\right) \,\Theta(\upvarpi-m_f).
 	\end{split}
 \end{align}
 The above result states that this interaction also contributes to the dipole part of GW radiation.
 
 \subsection{Total GW radiation in GBD theory}
 
 Now, considering the results obtained in the previous parts, i.e. Eqs. \eqref{Etensor}, \eqref{des11}, \eqref{Edotphi1} and \eqref{Edotcoup}, one can get the  total GW radiation in GBD theory as follows
 \begin{eqnarray}\label{totalE}
 	\dot{E}_{\mathrm{GW}}&=&\,G(1-\xi)\tilde{\mathfrak{g}}^{4/3}(M_c\,\upvarpi)^{10/3}\Bigg(-\,\frac{32}{15}\,+\,\frac{8}{9}(m\upvarpi\,\tilde{\mathfrak{g}})^{-2/3}\mathcal{S}^2\,\Bigg[\big(1-\frac{m_f^2}{\upvarpi^2}\big)^{3/2}\Theta(\upvarpi-m_f)\nonumber\\
 	&&-\frac{3(2\omega_0+1)^2}{8\omega_{0}^2(2\omega_{0}+3)}+\frac{2\omega_{0}+1}{4\omega_{0}^2(2\omega_{0}+3)}\big(1-\frac{m_f^2}{\upvarpi^2}\big)^{1/2}\cos (\mathfrak{R}\frac{m_{f}^2}{\upvarpi})\,\Theta(\upvarpi-m_f)\Bigg]\Bigg).
 \end{eqnarray}
 In the next section, we will imply the above result to draw exclusion plots in the two-dimensional parameter space of the theory, $ (m_f, \omega_{0})$, using observational data for the period derivative of compact binaries.

 \section{Constraints on GBD theory from $\dot{T}$ in compact binaries }
 In order to put some constraints on the parameter space $(\omega_{\rm BD},m_f)$ using the period derivative of compact binaries, an important step is to choose an appropriate binary system. Due to the presence of the difference in sensitivities ($\mathcal{S}=s_1-s_2$)
 in the scalar field contribution to the total GW radiation
 \eqref{totalE}, the best candidate for illustrating
 exclusion plots in the $(\omega_{\rm BD},m_f)$ plane are mixed
 binaries. In this regard, white dwarf-neutron star (WD-NS) binaries are particularly
 suitable owing to the large difference in sensitivities ($\sim10^{-4}$
 and $\sim0.2$ for WDs and NSs, respectively \cite{Zaglauer:1992bp}). One of these binaries for which accurate measurements of $\dot{T}$ and other necessary parameters have been taken is known as PSR J1012+5307, which is a 5.3 ms pulsar in a 14.5 hr quasicircular binary system with a low-mass WD companion \cite{PSR}.
 We have provided a summary of its properties in Table \ref{PSR J1012+5307}.
 \begin{center}
 	\begin{table}[htb]
 		\centering
 		\caption{Parameters relevant to the binary system PSR J1012+5307 \cite{PSR1}.}
 		\label{PSR J1012+5307}
 		\begin{tabular}{l r}
 			\hline
 			\hline
 			Period, $P$ (days) & 0.60467271355(3) \\
 			Period derivative (observed), $\dot{P}^{\rm obs}$ & $5.0(1.4)\;10^{-14}$ \\
 			Period derivative (intrinsic), $\dot{P}^{\rm intr}$ & $-1.5(1.5)\;10^{-14}$ \\
 			Mass ratio, q & 10.5(5) \\
 			NS Mass, $m_1$ ($M_\odot$) & 1.64(22) \\
 			WD Mass, $m_2$ ($M_\odot$) & 0.16(2) \\
 			Eccentricity, $e$ ($10^{-6}$) & 1.2(3) \\
 			\hline
 			\hline
 		\end{tabular}
 	\end{table}
 \end{center}
 The general approach to getting bounds on $(\omega_{\rm BD},m_f)$ using
 observations of the period derivative of mixed binaries is as
 follows: In the first step, the relation $(\dot{T}/T)=-\frac{3}{2}(\dot{E}/E)$ as well as
 $E=T+V=-\frac{1}{2}\mu v^2$, where  $v=(\tilde{\mathfrak{g}}m\,\upvarpi)^{1/3}$, should be used to calculate the fractional period decay due to the emission of GW radiation in GBD theory which is obtained as follows
 \begin{eqnarray}\label{T ratio}
 	\label{fractional T}
 	\frac{\dot{T}}{T}&=&-2 \,G(1-\xi)\mu\upvarpi^{2}\Bigg(\frac{4}{5}\,\,(\tilde{\mathfrak{g}}m\upvarpi)^{2/3}-\,\frac{1}{6}\mathcal{S}^2\,\Bigg[\big(1-\frac{m_f^2}{\upvarpi^2}\big)^{3/2}\Theta(\upvarpi-m_f)\nonumber\\
 	&&-\frac{3(2\omega_0+1)^2}{8\omega_{0}^2(2\omega_{0}+3)}+\frac{2\omega_{0}+1}{4\omega_{0}^2(2\omega_{0}+3)}\big(1-\frac{m_f^2}{\upvarpi^2}\big)^{1/2}\cos (\mathfrak{R}\frac{m_{f}^2}{\upvarpi})\,\Theta(\upvarpi-m_f)\Bigg]\Bigg)\,.
 \end{eqnarray}
 Now, we need to write the above relation in terms of the observables related to the system under inspection, which for circular binaries, the relevant observables are the stellar masses (including the mass ratio $q$) and the period. Rewriting Eq. \eqref{T ratio} in terms of these observables leads to the following relation
 \begin{eqnarray}\label{T dot}
 	\dot{T}&=&\dot{T}_\mathrm{GR}\,G^{-2/3}(1-\xi)\tilde{\mathfrak{g}}^{2/3}\Bigg(\frac{1}{12}\,-\,\frac{5}{144}\bigg(\frac{2\,\pi\,m\,\tilde{\mathfrak{g}}}{T}\bigg)^{-2/3}\mathcal{S}^2\,\Bigg[\big(1-\frac{T^2\,m_f^2}{4\pi^2}\big)^{3/2}\Theta(\upvarpi-m_f)\nonumber\\
 	&&-\frac{3(2\omega_0+1)^2}{8\omega_{0}^2(2\omega_{0}+3)}+\frac{2\omega_{0}+1}{4\omega_{0}^2(2\omega_{0}+3)}\big(1-\frac{T^2\,m_f^2}{4\pi^2}\big)^{1/2}\cos (\frac{\mathfrak{R}\,T\,m_{f}^2}{2\pi})\,\Theta(\upvarpi-m_f)\Bigg]\Bigg)
 \end{eqnarray}
 where the dimensionless quantity $\dot{T}_\mathrm{GR}$ denotes the GR prediction of period derivative of the system, given by
 \begin{align}
 	\label{pdotgrcirc}
 	\dot{T}_\mathrm{GR}=-\frac{192\pi}{5}\frac{q}{(1+q)^2}\left(\frac{2\pi G m}{T}\right)^\frac{5}{3}.
 \end{align}
 and using the parameters listed in Table \ref{PSR J1012+5307}, its value is estimated as $\dot{T}_\mathrm{GR}\approx -1.1\times 10^{-14}$.\par
 
 With the predicted $\dot{T}$ expressed in terms of the relevant set of parameters,
 we are in a position to compare it to observations; To this end, we use the fact that the value range of the predicted $\dot{T}$ should be less than the values of observed $\dot{T}_\mathrm{obs}$. It is notable that due to the presence of the Heaviside function in Eq. \eqref{T dot} and considering the period of the binary system PSR J1012+5307, our results are limited to the cases in which the mass of the scalar field is in the interval $m_{f}<8\times 10^{-29} \text{GeV}$. Moreover, according to the value reported for the period of this binary and the relation $\upvarpi^2=\frac{GM}{r^3}$, the relative distance of two binaries is of the order $r\approx 10^{24}\, \text{GeV}^{-1}$. This result leads to the conclusion that $m_{f}\,r\ll 1$, and therefore, we will work in this approximation henceforth. \par
 
\begin{figure}[tb]
 	\center
 	{   \includegraphics[scale=0.8]{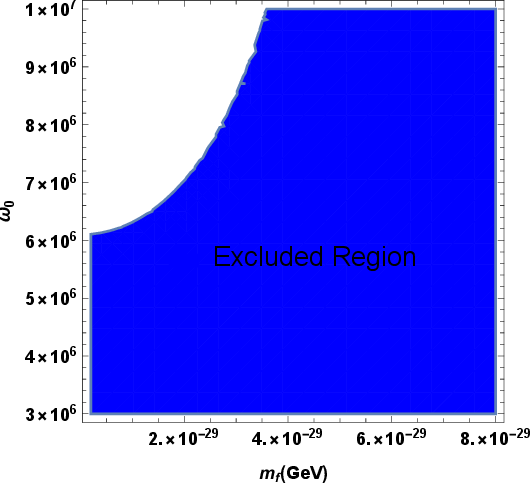} \hspace*{1cm}
 		\includegraphics[scale=0.8]{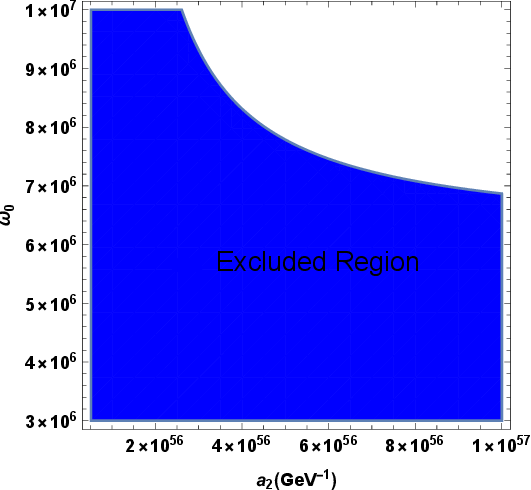}}
 	\caption{ Lower bound on $\omega_{0}$ as a function of
 		the mass of the scalar $m_f$ (left panel) and $a_2$ (right panel) from PSR J1012+5307.  Note that we set $G=6.7\times 10^{-39}\rm{GeV^{-2}}$.} \label{fig1}
 \end{figure}

 Considering all the mentioned points \textcolor{black}{as well as $\omega_0 \gg 1$}, the bound on $\omega_{0}$ will be obtained as a function of the geometrical
 scalar mass which is displayed in the left panel of Fig. \ref{fig1}. In particular, we find a lower bound $\omega_{0}>6.09723\times10^6$ for $m_{f}<8\times 10^{-29} \text{GeV}$. \textcolor{black}{ Here, it is required to mention that according to the  geometrical scalar field's mass relation, $m_{f}^{2}=\frac{2\omega_{0}+3}{6\omega_{0} a_{2}}$, there is a degeneracy between $\omega_0$ and $a_2$. However, working on the approximation $\omega_0 \gg 1$ leads to disappearance of the degeneracy and hence $m_f\simeq \frac{1}{3 a_2}$. Making use of this point,  we can plot the parameter space of $(\omega_0, a_2)$ for $a_2 > 5.2\times 10^{55} \text{GeV}^{-1}$ (right panel of Fig. \ref{fig1}).}

 \color{black}
 Before ending this part, it is worth comparing our result with the
 constraints on $\omega_0$ obtained from Solar System data. Indeed,
 Cassini measurements of the Shapiro time delay in the Solar System
 yield a lower bound $\omega_0 > 40000$ for the BD scalar masses
 smaller than $2.5\times 10^{-20} \text{eV}$ \cite{Alsing:2011er}.
 This is while, our calculations show a lower bound which is around
 two orders of magnitude more stringent than the bounds provided by
 the solar system data, for a massless BD scalar field and the
 geometrical field whose mass is smaller than $10^{-29}
 \text{GeV}$. Our finding indicates that the constraint obtained
 for the $\omega_0$ from compact binary systems is closer to the
 General Relativity values $\omega_0 \to \infty$.

 \color{black}
 \section{Frequency domain GW phase function }\label{phase function}
 We proceed this section by following the procedures described in \cite{Liu:2020moh} to work out the phase shift that GW suffers in the frequency domain during its propagation. Going
 to the frequency domain, we get the following relation for the GW waveform
 \begin{equation}
 	h(f)=\int h(t)e^{i2\pi ft}dt.
 \end{equation}
 We can solve the Fourier transform analytically using the stationary phase approximation for each mode which results in the following relation
 \begin{equation}
 	h(f)=\mathcal{A}(f)e^{i\Psi(f)}.
 \end{equation}
 It is important to note that all the relevant factors related to the polarization of the wave, distance to the source, and its position on the sky are included in the amplitude $\mathcal{A}$. Moreover, $\Psi(f)$ denotes the phase function of the GW which is given by \cite{Liu:2020moh}
 \begin{align}\label{PsiP3}
 	\begin{split}
 		\Psi(f)=&2\pi f(\mathfrak{R}+t_c)-2\Phi_c-\frac{\pi}{4}+\int_\infty^{\pi f}\frac{2\pi f-2 \upvarpi}{\dot{\upvarpi}}d\upvarpi
 	\end{split}
 \end{align}
 where $\Phi_{c}$ is formally defined as the phase of the wave at the time of coalescence, $t_c$ and $\upvarpi(t_c)=\infty$. As the above relation shows, in order to obtain the frequency-domain GW phase function, we need the time evolution of the orbital frequency $\upvarpi$. In order to achieve this aim, one can use the energy balance condition $\frac{dE}{dt}=-\frac{dE_{GW}}{dt}$ with $E=-\frac12\mu v^2=-\frac12\mu(\tilde{g}m\omega)^{\frac23}$ to find the general relation for
 the time derivative of the orbital frequency
 \begin{equation}
 	\dot{\upvarpi}=\frac{3\dot{E}}{\tilde{\mathfrak{g}}^{2/3}M_{c}^{5/3}}\upvarpi^{1/3}
 \end{equation}
 Making use of the results obtained in section \ref{GW radiation}, i.e. Eqs. \eqref{Etensor}, \eqref{des11}, \eqref{Edotphi1} and \eqref{Edotcoup} as well as working in the approximation that $\upvarpi\gg m_{f}$ and $m_f$ is a heavy scalar mass, the time evolution of the orbital frequency will be obtained as follows
 \begin{align}
 	\begin{split}
 		\dot{\omega}=&-\frac{32}{15}\beta^{2/3}\Big(GM_{c}(1-\xi)\Big)^{5/3}\upvarpi^{11/3}\Bigg[1-\frac{40}{32}\mathcal{S}^2\Big(G(1-\xi)\beta m \upvarpi\Big)^{-2/3}\\
 		&\Bigg\{(1-\frac{3m_{f}^2}{2\upvarpi^2})-\frac{3(2\omega_{0}+1)^2}{8\,\omega_{0}^{2}(2\omega_{0}+3)}\Bigg\}\Bigg],
 	\end{split}
 \end{align}
 where
 \begin{align}
 	\beta=1+\frac{1}{(2\omega_{0}+3)}\,(1-2s_1)(1-2s_2).
 \end{align}
 Now, we are ready to evaluate the integration in Eq. \eqref{PsiP3} which becomes
 \begin{align}
 	\begin{split}
 		&\int_\infty^{\pi f}\frac{2\pi f-2 \upvarpi}{\dot{\upvarpi}}d\upvarpi\\
 		=&\int_\infty^{\pi f}d\upvarpi(2\pi f-2\upvarpi)(\frac{-15}{32})\Big(GM_c(1-\xi)\Big)^{-5/3}\beta^{-2/3}\upvarpi^{-{11}/{3}}\Bigg[1+\frac{40}{32}\mathcal{S}^2\Big(G(1-\xi)\beta m \upvarpi\Big)^{-2/3}\\
 		&\Bigg\{(1-\frac{3m_{f}^2}{2\upvarpi^2})-\frac{3(2\omega_{0}+1)^2}{8\,\omega_{0}^{2}(2\omega_{0}+3)}\Bigg\}\Bigg]\\
 		=&-\frac{72}{128}\beta^{-2/3}(GM_c\pi f)^{-5/3}\Bigg(1+\frac{5}{3}\xi+\frac{25}{9}\frac{\mathcal{S}^2}{(G\beta m \pi f)^{\frac23}}(1+\frac{7}{3}\xi)\\
 		&\Bigg[\frac{9}{35}+\frac{53}{208}\Big(\frac{m_{f}}{\pi f}\Big)^2+\frac{27}{280}\frac{(2\omega_{0}+1)^2}{\,\omega_{0}^{2}(2\omega_{0}+3)}\Bigg]\Bigg).
 	\end{split}
 \end{align}
 Since for a light BD scalar mass ($m_s|_{BD}<2.5\times10^{-20}~\text{eV}$), the Cassini spacecraft has constrained $\omega_0$ to be larger than 40~000 \cite{Alsing:2011er}, i.e. $\xi<10^{-5}$, we retain only terms to the linear order of $\xi$ in the above equation. Then, the phase $\Psi$ in the massless BD scalar field situation becomes
 \begin{align}\label{PsiP4}
 	\begin{split}
 		\Psi=&2\pi f(\mathfrak{R}+t_c)-2\Phi_c-\frac{\pi}{4}-\frac{72}{128}\beta^{-\frac23}(GM_c\pi f)^{-\frac53}\Bigg(1+\frac{5}{3}\xi+\frac{25}{9}\frac{\mathcal{S}^2}{(G\beta m \pi f)^{\frac23}}(1+\frac{7}{3}\xi)\\
 		&\Bigg[\frac{9}{35}+\frac{53}{208}\Big(\frac{m_{f}}{\pi f}\Big)^2+\frac{27}{280}\frac{(2\omega_{0}+1)^2}{\,\omega_{0}^{2}(2\omega_{0}+3)}\Bigg]\Bigg),
 	\end{split}
 \end{align}
 which clearly shows how the parameters involved in GBD theory can affect the phase function of the GWs.
 
 \section{ Conclusion}\label{conclusion}
 In this paper, we set constraints on the GBD theory, which is a
 modified BD theory by generalizing the Ricci scalar $R$ in the
 original BD action to an arbitrary function $f(R)$, with an action
 of the form \eqref{action}, assuming that the BD scalar field is
 massless and the coupling function $\omega$ is constant.
 Specifically, we calculated the derivative of the orbital period
 for quasicircular binaries. The calculation steps of which are as
 follows:
 
 We started by studying the weak field equations in GBD theory. Our
 calculation showed that the GBD theory could be considered as two
 scalar-fields theory, i.e. the BD field and the effective
 geometrical field $f_{R}=\Phi$. Next, we solved the obtained
 linearized field equations to Newtonian order for point particles
 and used the EIH equations of motion to describe the approximate
 dynamics of the system. Since we aimed to calculate the
 gravitational radiation power of all propagating degrees of
 freedom in GBD theory, we made use of the perturbed field
 equations method to find the GW SET. In this regard, detailed
 calculations were provided for the GW radiation power originating
 from both tensor field and scalar fields. Based on our
 calculation, both scalar fields contribute to GW radiation by
 producing dipole radiation. In the continuation, the obtained
 results were utilized to calculate the period derivative of
 compact binaries in GBD theory. Then, we used the observational
 data of PSR J1012+5307 binary system to put bounds on some
 parameters of the desired theory by sketching the relevant
 parameter space shown in Fig. \eqref{fig1}. Our findings indicate
 a lower bound of $\omega_{0}>6.09723\times10^6$ for the mass range
 $m_{f}<8\times 10^{-29} \text{GeV}$. \textcolor{black}{It is worth
 	mentioning that the obtained constraint for $\omega_0$ is two
 	orders of magnitude more stringent than the constraints provided
 	by the solar system data, demonstrating that our result for the
 	$\omega_0$ constraint is closer to the General Relativity values
 	$\omega_0 \to \infty$. Studying the bound on the $\omega_0$ as a
 	function of the mass of geometrical scalar field $(m_f)$ coming
 	from Solar System data will be addressed in an independent work.}
 \par
 
 Moreover, we found the phase shift that GWs experience in the
 frequency domain during their propagation which is predicted by
 GBD theory.\par
 
 According to our analysis, it is worth investigating the
 applicability of the parameterized post-Einsteinian (ppE)
 framework \cite{Yunes:2009ke} in terms of the parameters of the
 theory. Besides, one can generalize the obtained solutions in the
 presence of the massive BD scalar field as well as $\omega$ being
 a function of the massive scalar field. \textcolor{black}{
 	Moreover, this analysis opens the door to exploring similar
 	applications in other modified BD gravity frameworks
 	\cite{Kofinas:2016fcp} to study their theoretical parameters more
 	precisely.} These topics will be addressed in independent works.

 \section{Acknowledgment}
 The authors are grateful to the anonymous reviewer for useful
 comments. This work is based upon research funded by Iran National
 Science Foundation (INSF) under project No. 4025951.


\end{document}